\documentclass{vldb}
\usepackage{pgfplotstable}
\usepackage{booktabs} 
\usepackage{siunitx}
\usepackage{color}
\usepackage{tabularx}
\usepackage{multirow}
\usepackage{balance}
\usepackage{scrextend}
\usepackage{listings}
\usepackage{subcaption}
\usepackage{tikz}
\usetikzlibrary{positioning}
\usepackage{pifont}
\usepackage{url,xspace}
\usepackage{algorithm}
\usepackage{amsmath}
\usepackage{bigstrut,cite}
\usepackage{float}
\usepackage[bookmarks=false]{hyperref}
\usepackage{cleveref}
\usepackage{bold-extra}
\usepackage{graphicx}

\usetikzlibrary{shapes,arrows}

\newcolumntype{P}[1]{>{\centering\arraybackslash}p{#1}}

\makeatletter
\renewcommand{\paragraph}{%
    \vskip 8pt\@startsection{paragraph}{3}{\z@}{6\p@ \@plus \p@}%
    {-5\p@}{\subsecfnt}%
}
\newcommand{\paragraphafterexample}{%
    \vskip 0pt\@startsection{paragraph}{0}{\z@}{0\p@ \@plus \p@}%
    {-5\p@}{\subsecfnt}%
}
\makeatother

\newcommand{\EHP}{LevelHeaded\xspace}
\newcommand{\EH}{EmptyHeaded\xspace}

\newcommand{\rv}[1]{{\color{black}#1}}
\newcommand{\numb}[1]{{\color{black}#1}}

\def\compactify{\itemsep=0pt \topsep=0pt \partopsep=0pt \parsep=0pt}

\definecolor{linenumbercolor}{rgb}{0.5,0.5,0.5}
\newdef{definition}{Definition}
\newtheorem{example}{Example}[section]
\newtheorem{principle}{Crucial Observation}[section]

\newfloat{algorithm}{t}{lop}

\title{LevelHeaded: Making Worst-Case Optimal Joins\\Work in the Common Case}

\numberofauthors{1}
\author{
\alignauthor Christopher R. Aberger, Andrew Lamb, Kunle Olukotun, and Christopher R\'e\\
  \affaddr{Stanford University}\\
  \email{\{caberger,lamb,kunle,chrismre\}@stanford.edu}
}

\begin{document}
\sloppy
\pagestyle{empty}

\maketitle
\begin{abstract} 
Pipelines combining SQL-style business intelligence (BI) queries
and linear algebra (LA) are becoming increasingly common in industry. As a result,
there
is a growing need to unify these workloads in a single
framework. Unfortunately, existing solutions either sacrifice the inherent
benefits of exclusively using a relational database (e.g. logical
and physical independence) or
incur orders of magnitude performance gaps compared to specialized engines (or both).
In this work we study applying a new type of query processing architecture to
standard BI and LA benchmarks.  To do this we present a new in-memory query 
processing engine called \EHP. \EHP uses worst-case optimal joins as its core
execution mechanism for both BI and LA queries.
With \EHP, we show how crucial optimizations for
BI and LA queries can be captured in
a worst-case optimal query architecture. Using these optimizations, \EHP 
outperforms other relational database engines (LogicBlox,
MonetDB, and HyPer) by \numb{orders of magnitude} on standard LA 
benchmarks, while performing on average within 31\%
of the best-of-breed BI (HyPer) and LA (Intel MKL) solutions on their own
benchmarks. Our results show that such a \emph{single} query processing architecture is
capable of delivering competitive performance on \emph{both} BI and
LA queries.
\end{abstract}

\vspace{-0.5mm}
\section{Introduction}
The efficient processing of classic SQL-style workloads is no longer enough;
machine learning algorithms are being adopted at an explosive rate. In fact,
Intel projects that by 2020 the hardware cycles dedicated to machine
learning tasks will grow by 12x, resulting in more servers running
this than any other workload \cite{intelbuzz}. As a result, there is a
growing need
for query processing engines that
are efficient on (1) the SQL-style queries at the core of most
business intelligence workloads and (2) the linear algebra operations
at the core
of most machine learning algorithms. In this work, we explore whether a new
query processing architecture is capable of delivering competitive
performance in both cases.

An increasingly popular workflow combines business intelligence (BI) and linear algebra (LA) queries  
by executing SQL queries in a relational warehouse as a means to extract 
feature sets for machine learning models. 
Unsurprisingly, these SQL queries are similar to standard BI workloads: 
the data is de-normalized (via joins), filtered, and aggregated to form a single feature set 
\cite{kumar2016join,kumar2016model}. Still, BI queries 
are best processed in a relational database
management system (RDBMS) and LA queries are best 
processed in a
LA package. As a
result, there has been a flurry of activity around building systems
capable of unifying both BI and LA
querying \cite{hellerstein2012madlib,brown2010overview,luo2017scalable,armbrust2015spark,kernert2015bringing,utnla,duggan2015bigdawg}. 
At a high-level, existing approaches fall into one of three classes:
\vspace{-1.5mm}
\begin{itemize}
\setlength\itemsep{0.01em}
\item \emph{Exclusively using a relational engine.} There are many
inherent advantages to exclusively using a RDBMS to process both BI and LA queries.
Simplifying extract-transform-load 
(ETL), increasing usability, and leveraging well-known
optimizations 
 are just a few \cite{luo2017scalable}. Although it is known that 
LA queries can be expressed using joins and aggregations, executing these queries 
via the pairwise join algorithms in standard RDBMSs is orders of magnitude slower 
than using a LA package (see \Cref{sec:experiments}). 
Thus, others \cite{luo2017scalable} have shown that a RDBMS must be modified
to compete on LA queries. 
\item \emph{Extending a linear algebra package.} Linear algebra packages, like 
BLAS \cite{blackford2002updated}
or LAPACK \cite{anderson1999lapack}, provide high-performance through low-level 
procedural interfaces and therefore lack the ability for high-level querying. To 
address this, array databases with high-level querying, like SciDB 
\cite{brown2010overview}, have been proposed. Unfortunately, array databases are 
highly specialized and are not designed for general BI querying. 
As a result, support for SQL-style BI querying 
\cite{armbrust2015spark,mckinney2011pandas} has recently been
combined with the LA support found in popular packages 
like Spark's MLlib \cite{meng2016mllib} and Scikit-learn 
\cite{pedregosa2011scikit}. 
Still, these solutions lack many of the inherent benefits of a RDMBS, 
like a sophisticated (shared-memory) query optimizer or efficient
data structures (e.g. indexes) for query execution, and therefore achieve suboptimal 
performance on BI workloads (see \Cref{sec:extensions}).
\item \emph{Combining a relational engine with a linear algebra package.}
To preserve the benefits of using a RDBMS on BI queries, while also
avoiding their pairwise join algorithms on LA queries, 
others (e.g. Oracle's UT\_NLA \cite{utnla} and MonetDB/NumPy \cite{monetdbnumpy})
have integrated a RDBMS with a LA package. 
Still, these approaches
just tack on an external LA package to a RDBMS---they do not fully
integrate it. Therefore, users are forced to write low-level code wrapping
the LA package and the LA computation is a black box to
the query optimizer. Even worse, this integration, while straightforward on
dense data, is complicated (and largely unsolved) on sparse data because a RDBMS 
and a LA package use fundamentally different 
data layouts here (e.g. column store versus
compressed sparse row).
\end{itemize}
\vspace{-1mm}
Therefore, existing approaches either (1) sacrifice the inherent benefits of using
only an
RDBMS to process both classes of queries, (2) are unable to process both classes
of
queries, or (3) incur orders of magnitude performance gaps relative to the best 
single-class approach.

\begin{figure}
 \centering
 \includegraphics[width=0.76\linewidth]{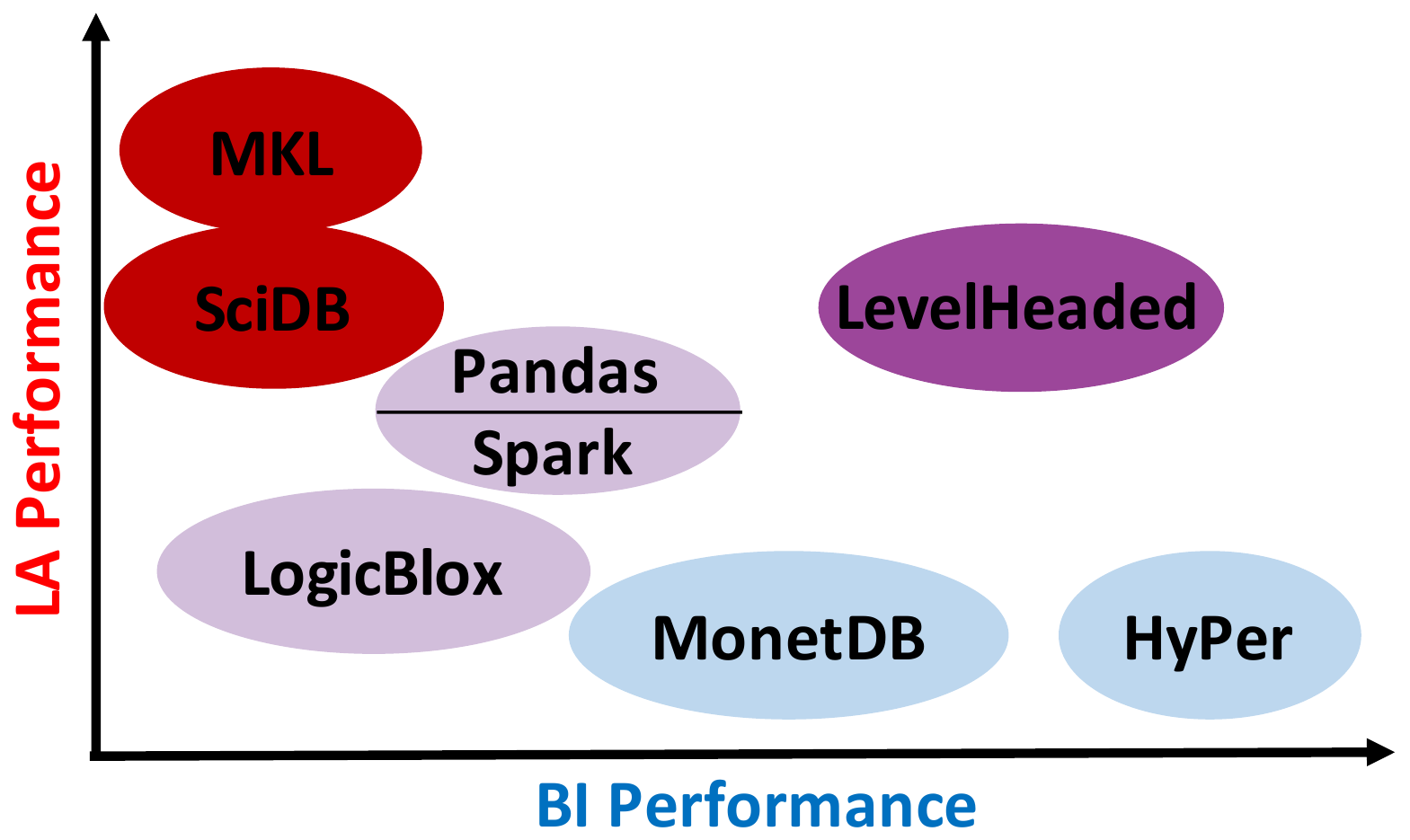}
\caption{The relative performance of popular
engines and target performance for \EHP on
business intelligence (BI) and linear algebra (LA) queries.}
\label{fig:systems_chart}
\vspace{-2mm}
\end{figure}

In this work we study an alternative approach to building a RDBMS
for both BI and LA querying. In particular, we
study using worst-case optimal join (WCOJ) algorithms \cite{ngo2012worst} as the
mechanism to unify these query workloads. To do this, we present a new in-memory query 
processing engine called \EHP. \EHP uses a novel WCOJ query architecture to optimize,
plan,  and 
execute both BI and LA queries. In contrast 
to previous WCOJ engines \cite{aberger2015emptyheaded}, 
\EHP is designed for and evaluated on more than just graph
queries. As such, \EHP is the first WCOJ engine to present an evaluation on 
both BI and LA queries. In contrast to other query engines 
\cite{brown2010overview,zaharia2010spark,luo2017scalable,idreos2012monetdb,kemper2011hyper,aref2015design}, 
\EHP competes with both LA packages and RDBMSs
on their own benchmarks (see \Cref
{fig:systems_chart}).

However, designing a new query processing engine that
is efficient on both BI and LA queries is a
challenging task. The recently proposed WCOJs at the
core of this
new query processing architecture are most effective on graph queries 
where they have an asymptotic advantage over traditional
pairwise join algorithms.
In contrast, pairwise
join algorithms are well-understood, and have the
advantage of
45+ years of proven constant factor optimizations for BI
workloads
\cite{stonebraker2005c,raman2013db2,shute2013f1,delaney2000inside,kemper2011hyper,idreos2012monetdb}.
Further, LA queries, which also have the benefit of
decades of optimizations 
\cite{whaley1998automatically,anderson1999lapack,blackford2002updated}, are
typically
not a good match for the relational model. Therefore, it is not at all
obvious whether they are good match for this new type of relational query
processing architecture.

\begin{figure}
 \centering
 \includegraphics[width=8.5cm]{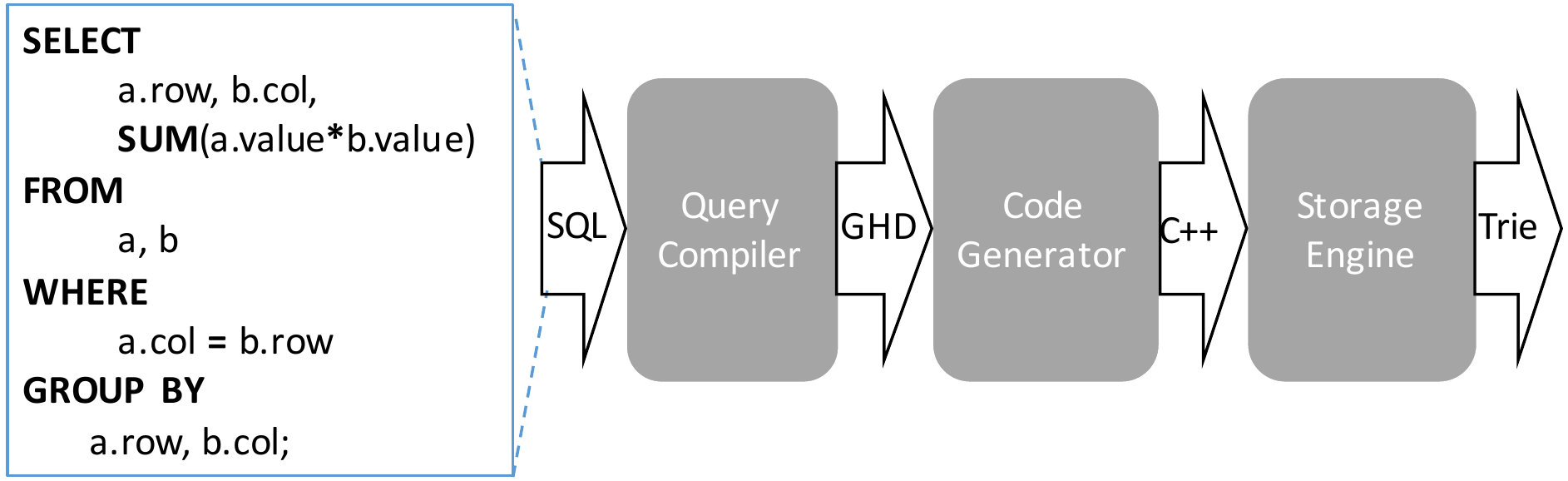}
\caption{System overview with matrix multiplication input.}
\label{fig:system_overview}
\end{figure}

Despite these challenges, we found that the unification of three new
techniques in a WCOJ architecture could enable it 
 to deliver competitive performance on both 
BI and LA queries. The techniques we leverage are (1) a new mechanism to translate general
SQL queries to WCOJ query plans, (2) a new cost-based query optimizer for 
WCOJ's, and (3) a simple storage engine optimizer for \texttt{GROUP BY}. It was
not at all obvious how to identify and unify these techniques in a WCOJ framework and,
as such, we are the first to do so. The three core techniques of \EHP
in more detail are:
\vspace{-4mm}
\begin{itemize} \compactify
\setlength\itemsep{0.01em}
\item \emph{SQL to GHDs: Pushing Down Selections and Attribute
Elimination.} Neither the theoretical literature \cite{ngo2014skew} nor the query compilation
techniques for WCOJs \cite{joglekar2016ajar} maps directly to all
SQL features. In \EHP we implement a practical extension of these
techniques that enables us to capture more general query workloads
as well as the classic query optimizations of pushing down selections and attribute 
elimination. Besides providing up to \numb{4x} speedup on BI 
queries, a core artifact of our attribute elimination implementation is that it 
enables \EHP to target BLAS packages on dense LA queries
at little to no execution cost. This is because it is challenging to outperform BLAS 
packages, like Intel MKL \cite{intelmkl} on sparse data and is usually not 
possible\footnote{\small Intel MKL often gets peak hardware efficiency on dense 
LA \cite{intelmkl}.} on dense data. Therefore, 
\EHP leverages attribute elimination to opaquely call Intel MKL on dense LA 
queries while executing sparse LA queries as pure aggregate-join queries (entirely in 
\EHP).
\item \emph{Cost-Based Optimizer: Attribute Ordering.}
WCOJ query optimizers need to select an attribute order \cite{ngo2012worst} in 
a similar manner to how traditional query optimizers select a join order 
\cite{garcia2008database}. With \EHP, we present a 
cost-based optimizer to select a WCOJ attribute order for the 
first time. We highlight that our optimizer follows heuristics
that are different from what conventional wisdom from pairwise join optimizers 
suggests (i.e. highest cardinality first) and describe how to leverage these 
heuristics to provide an accurate cost-estimate for a WCOJ algorithm.  
Using this cost-estimate, we validate that \EHP's 
cost-based optimizer selects attribute orders than can be up to
\numb{8815x} faster than attribute orders that previous WCOJ engines 
\cite{aberger2015emptyheaded} might select.
\item \emph{Group By Tradeoffs: Battling Skew on Group By.} Finally, 
we study the classic tradeoffs around executing \texttt{GROUP BY}s  
in the presence of skew. Although these tradeoffs
are well understood for BI queries, they have yet to be applied to this new
style of query processing engine and are also essential for LA queries 
(see \Cref{fig:system_overview}). We describe \EHP's \texttt{GROUP BY} 
optimizer that exploits these tradeoffs and show that it provides
up to a \numb{875x} and \numb{185x} speedup over a naive implementation on BI 
and LA queries respectively.
\end{itemize}
\vspace{-1mm}

We evaluate \EHP on standard BI and
LA benchmarks: seven TPC-H queries\footnote{\small The TPC-H queries
are run without the \texttt{ORDER BY} clause.} and four (two sparse, two dense) 
LA kernels. These benchmarks are de-facto standards
for relational query processing and LA engines. Thus, each engine
we compare to is designed to process one of these benchmarks efficiently
using specific optimizations that enable high-performance on one type 
of benchmark, but not necessarily the other.
Therefore, although these engines are the state-of-the-art solutions within a
benchmark, they are unable to remain competitive across benchmarks.
For example, HyPeR delivers high performance on BI queries, but is not 
competitive on most LA workloads; similarly, Intel's Math Kernel Library (MKL) 
\cite{intelmkl} delivers high performance on LA queries, but does not provide 
support for BI querying. In contrast, \EHP is
designed to be generic, maintaining efficiency across the queries in both 
benchmarks.

\paragraph*{Contribution Summary } This paper introduces the \EHP engine and
demonstrates that its novel architecture can compete on standard BI and LA 
benchmarks. We show
that \EHP outperforms other relational engines by at least \numb{an order of
magnitude}
on LA queries, while remaining on average within \numb{31\%} of 
best-of-the-breed solutions on \emph{both} BI and LA queries.

Our contributions and outline are as follows. 
\begin{itemize} \compactify
	\item In \Cref{sec:preliminaries}, we describe the \EHP architecture. 
    In particular, we describe how 
\EHP's query and data model, which is different from that of previous WCOJ engines \cite{aberger2015emptyheaded,aref2015design}, preserves the theoretical benefits of a WCOJ architecture 
\emph{and} enables it to 
efficiently process BI and LA queries.
	\item In
\Cref{sec:query_optimization,sec:cost_based,sec:query_execution} we present the
core (logical and physical) optimizations that we unify in a
WCOJ query architecture for the first time. We show that these optimizations 
provide up to a \numb{three orders of magnitude} performance 
speedup and are necessary to compete on BI and LA queries. In more detail, we present the
classic query optimizations of pushing down selections
and attribute elimination in \Cref{sec:query_optimization}, a cost-based
optimizer for selecting an attribute order for a WCOJ algorithm in \Cref{sec:cost_based},
and, finally, a simple query execution optimizer for \texttt{GROUP BY}s in \Cref
{sec:query_execution}.    
\item In \Cref{sec:experiments} we show that \EHP can outperform other
relational
engines by \emph{an order of magnitude} on standard BI and LA benchmarks 
while remaining on average within \numb{31\%} of a best-of-breed solution within
each benchmark. For the first time, this evaluation validates that a WCOJ architecture can
compete on BI and LA workloads. We argue that the inherent benefits of 
such a unified (relational) design has the potential to outweigh its
minor performance overhead. \end{itemize}

We believe \EHP represents a first step in unifying relational algebra and
machine learning in a single engine. As such, we extend \EHP in \Cref
{sec:extensions} to explore some of the potential benefits of 
this approach on a full
application. We show here that \EHP is up to \numb{an order of magnitude} 
faster than the popular (unified) solutions of MonetDB/Scikit-learn, Pandas/Scikit-learn, 
and Spark on a workload that combines SQL-style querying and a machine
learning algorithm. We hope \EHP adds to the debate surrounding the
design of unified querying systems.

\subsection{Related Work}
\label{sec:related_work}

\EHP extends previous work in worst-case optimal join processing (\EH and
LogicBlox), relational data processing, and linear algebra processing.

\paragraph*{EmptyHeaded}
\EHP is designed around many of the techniques presented in the \EH engine 
\cite{aberger2015emptyheaded,aberger2016old}, but with many fundamental differences.
First, \EH was only evaluated in the graph and 
resource description framework (RDF) domains. As such, the \EH
design is unable to support BI
workloads that contain many attributes. The limitations in the \EH design stem
from its storage model and query model which was specifically designed for graph and
RDF workloads. In addition, the \EH design only
supported equality selections and \texttt{GROUP BY}'s on indexed attributes. 
Finally, the design of 
\EH was optimized for only sparse data and achieved suboptimal performance on
workloads with dense data. \EHP is designed to fix these flaws by leveraging
new techniques (\Cref{sec:data_model,sec:storage_model,sec:query_optimization,sec:cost_based,sec:query_execution})
that eliminate these limitations, 
while preserving the theoretical benefits of such a design.

\paragraph*{LogicBlox}

LogicBlox \cite{aref2015design} is a full featured commercial database engine built around similar
worst case optimal join \cite{veldhuizen2012leapfrog} and query compilation \cite{khamis2015sf} algorithms. Still, a systematic evaluation of the LogicBlox
engine on BI and LA workloads is yet to be presented. 
From our conversations with LogicBlox, we learned that they 
often avoid using a WCOJ algorithm on these workloads in favor of more traditional
approaches to join processing. In contrast, \EHP always uses a WCOJ algorithm  
(even on queries not approaching the worst-case). Further, LogicBlox uses a query
optimizer that has similar benefits to \EHP's (see 
\Cref{sec:query_model}), but does so using custom algorithms \cite{khamis2015sf}. 
In contrast, \EHP uses a generalization of these algorithms \cite{joglekar2016ajar} 
that maps to well-known techniques \cite{gottlob2005hypertree}. Finally,
LogicBlox stores its data in manner that resembles a row store, whereas
\EHP resembles column store.

\paragraph*{Relational Processing} A significant amount of work has focused on 
bringing LA to relational data processing engines. Some have
suggested treating LA objects as first class citizens in a column
store \cite{kernert2015bringing}. Others, such as Oracle's UTL\_NLA \cite{utnla}
and MonetDB with embedded Python \cite{monetdbnumpy} allows users to call LA 
packages through user defined functions. Still, the relational optimizers in these
approaches do not see, and therefore are incapable of optimizing, the 
LA routines. Even worse, these packages
are cumbersome to use and place significant burden on the user to make low-level
library calls. Finally, the SimSQL project \cite{luo2017scalable} suggests that relational
engines can be modified in straightforward ways to accommodate LA
workloads. Our goals are similar to the SimSQL, but explored with different
mechanics. SimSQL studied the necessary modifications for a classic
database architecture to support LA queries and
was only evaluated on LA queries. In \EHP, we evaluate a new,
theoretically advanced, query processing architecture on both BI and LA 
workloads. Additionally, SimSQL was evaluated
in the distributed setting and therefore compared to higher-level baselines. 
\EHP is evaluated in a shared memory setting against Intel MKL and HyPer 
\cite{kemper2011hyper}. Other high performance
in-memory databases, like HyPer, focused on classic OLTP and OLAP
workloads and were not designed with other workloads in mind.

\paragraph*{Linear Algebra Processing} Researchers have long studied how to
implement high-performance LA kernels. Intel 
MKL \cite{intelmkl} represents the culmination of this work on Intel CPUs. 
Unsurprisingly, researchers have shown  \cite{smith2014anatomy} that it requires
tedious low-level code optimizations to come near the performance of a package
such as Intel MKL. As a result, processing these queries in a traditional RDBMS 
(using relational operators) is at least an order of magnitude slower than using 
such packages (see \Cref{sec:experiments}). In response,
researchers have released array databases, like SciDB and TileDB
\cite{brown2010overview}, which provide high-level LA querying, often by 
wrapping lower-level libraries. 
In contrast, our goal is not to
design an entirely different and specialized engine for these workloads, but 
rather to design a single (relational) engine that processes multiple classes 
of queries efficiently.

\section{levelheaded Architecture}
\label{sec:preliminaries}

\begin{figure}
 \centering
 \includegraphics[height=2.5cm]{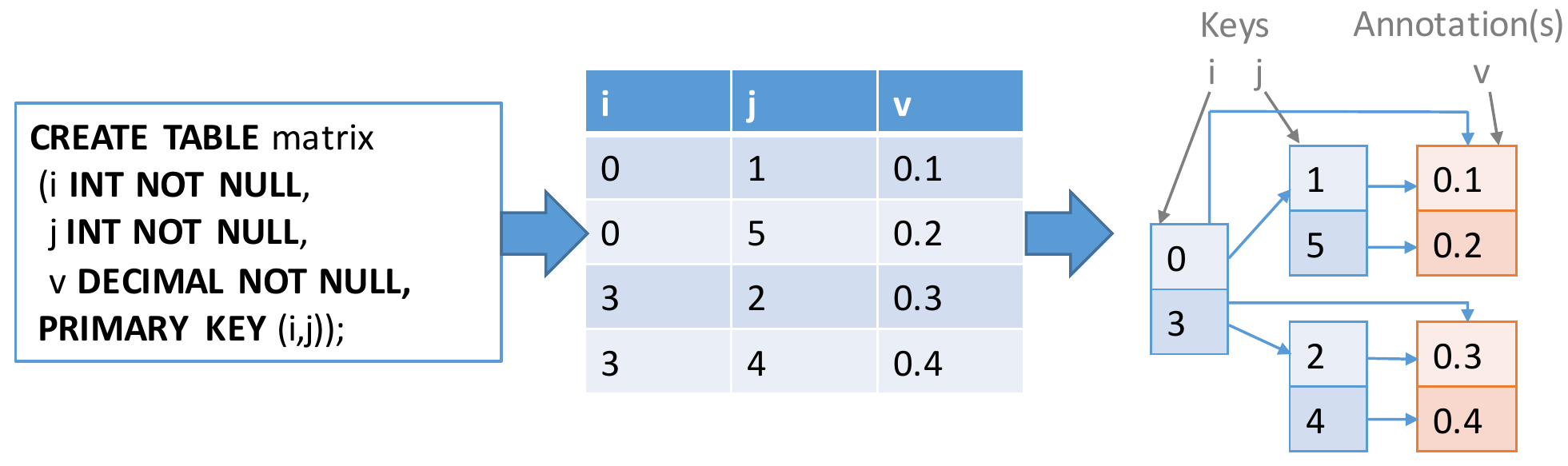}
\caption{Storage of a matrix in a \EHP trie.}
\label{fig:trie}
\end{figure}

\begin{figure*}
\centering
\includegraphics[width=0.90\linewidth]{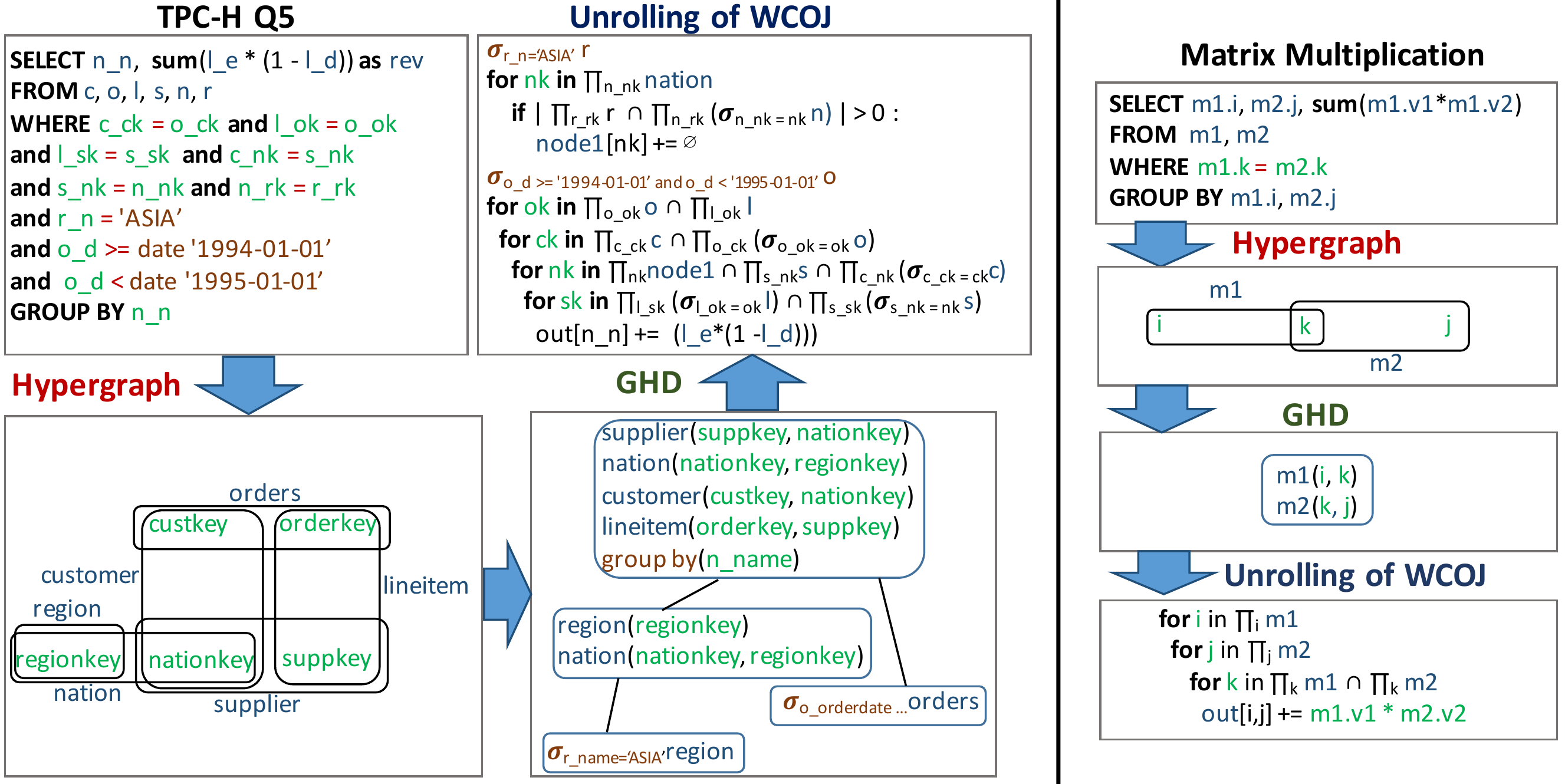}
\caption{Illustration of the core components of \EHP on TPC-H query 5 and matrix
multiplication. The
queries are expressed in SQL, translated to a
hypergraph, the hypergraph is used to generate an optimal GHD-based query plan,
and code instantiating the WCOJ algorithm is generated from
the resulting query plan. Relation names and attribute names are 
abbreviated (e.g. `nk' = `nationkey', `n' = `nation', ...) in the SQL and generated code for readability.}
\label{fig:core_components}
\end{figure*}

In this section we overview the data model, storage engine, query compiler, 
and join algorithm at
the core of the \EHP architecture. This overview presents the 
preliminaries necessary to understand the optimizations presented in \Cref{sec:query_optimization,sec:cost_based,sec:query_execution}. \EHP
ingests structured data from delimited
files on disk and has a Python front-end that accepts Pandas dataframes. 
The query language is a subset of the SQL 2008 syntax 
with some standard limitations that we
detail in \Cref{sec:data_model}. The input and output of each query is
a \EHP table which we describe in \Cref{sec:storage_model}. The SQL queries
are translated to generalized hypertree decompositions (GHD) which we describe
in
\Cref{sec:query_model}. Finally, code is generated from the selected GHD using
a WCOJ algorithm which we describe in \Cref
{sec:computation_model}. The entire process is shown in \Cref{fig:system_overview}. 

\subsection{Data Model}
\label{sec:data_model}
The \EHP data model is relational with some minor restrictions. A core
aspect of \EHP's data model is that attributes are classified as either \emph{keys}
or \emph{annotations} via a user-defined schema. Keys in \EHP correspond to primary or foreign
keys and are the only attributes which can partake in a join. Keys cannot be
aggregated. Annotations are all other attributes and can be aggregated. 
Both keys and annotations support filter predicates and
\texttt{GROUP BY} operations. In its current implementation, \EHP supports
equality (\texttt{=}) filters on keys and range (\texttt{>},\texttt{<},
\texttt{=}) filters on annotations. This represents the
existing implementation, not fundamental restrictions. \EHP's current
implementation supports
attributes with types of \texttt{int}, \texttt{long}, \texttt{float}, 
\texttt{double}, and \texttt{string}. In many
regards \EHP is similar to a key-value store combined with the relational
model. \EHP is not unique in this regard, and commercial databases like
Google's Mesa \cite{gupta2014mesa} and Spanner \cite{corbett2013spanner} follow 
a similar (if not identical) model.

\subsection{Storage Engine}
\label{sec:storage_model}
\EHP's data model is tightly coupled with how it stores relations. All key 
attributes from a relation are stored in a trie, which serves as the only 
physical index in \EHP. In the trie, each level is composed of sets of 
dictionary encoded (unsigned integer) values. As is 
standard \cite{aberger2015emptyheaded}, \EHP stores dense sets using 
a bitset and sparse sets using unsigned integers. Annotations are stored
in separate buffers attached to the trie. \EHP supports multiple
annotations, and each can be reached from any level of the trie (core 
differences from \EH). An 
example \EHP trie is shown in \Cref{fig:trie} where the dimensions 
\texttt{i} and 
\texttt{j} are keys and  
\texttt{v} is an annotation.
\EHP tries can be thought of as
an index on key attributes or a materialized view of the input table.

\subsection{Query Compiler}
\label{sec:query_model}

We briefly overview the theoretical foundation of \EHP's query compiler by
describing the essential details.

\paragraph*{Representing Queries as Hypergraphs}
Like \EH \cite{aberger2015emptyheaded}, a query is represented using a 
hypergraph $H = (V, E)$, where $V$ is a
set of vertices (attributes) and each hyperedge $e \in E$ 
(relation) is a subset of $V$. A join query is
represented as a subgraph of $H$. Example hypergraphs are shown in 
\Cref{fig:core_components}. In \Cref{sec:query_optimization} we describe
how SQL is translated to hypergraphs in more detail.

\paragraph*{Generalized Hypertree Decompositions}

The \EHP query compiler uses \emph{generalized hypertree decompositions} (GHDs) 
\cite{gottlob2005hypertree} to represent query plans. It is useful to think of 
GHDs as an analog of relational algebra for a WCOJ algorithm. Formally, given a 
hypergraph $H = (V_H, E_H)$, a GHD is a tree $T = (V_T, E_T)$ and a mapping 
$\chi : V_T \rightarrow 2^{V_H}$ that associates each node of the tree with a 
subset of vertices in the hypergraph.  The asymptotic
runtime of a plan is bound by the fractional hypertree width (FHW) of the GHD 
\cite{joglekar2016ajar} and, as is standard \cite{aberger2015emptyheaded},  
\EHP chooses a GHD with the lowest FHW to select a query plan that matches the best 
worst-case guarantee. 

\paragraph*{Capturing Aggregate-Join Queries}

Aggregate-join queries are common in both BI and LA workloads.
To capture aggregate-join queries, \EHP uses the AJAR framework 
\cite{joglekar2016ajar}. Despite that others \cite{khamis2015sf}
have argued that custom algorithms are necessary to capture such queries,
AJAR extends the theoretical results of 
GHDs to queries with aggregations. AJAR does this by associating each tuple 
in a one-to-one mapping with an \textit{annotation}. Aggregated annotations are 
members of a \textit{commutative semiring}, which is equipped with product and 
sum operators that satisfy a set of properties (identity and annihilation, 
associativity, commutativity, and distributivity). Therefore, when relations are joined, 
the annotations on each relation are multiplied together to form the annotation 
on the result. Aggregations are expressed by an \textit{aggregation ordering} 
$\alpha = (\alpha_1, \oplus_1), (\alpha_2, \oplus_2), \ldots$ of attributes and
operators. 

Using AJAR, \EHP picks a query plan with the best worst-case guarantee 
by going through three phases. 
First, it breaks the input query into characteristic
hypergraphs, which are subgraphs of the input that can be decomposed to optimal 
GHDs. Second, for each characteristic hypergraph all possible
decompositions are enumerated, and a decomposition that minimizes FHW is chosen.
Finally, the chosen GHDs are combined to form an optimal GHD. \numb{To avoid
unnecessary intermediate results, \EHP compresses 
all final GHDs with a FHW of 1 into a single 
node, as the query plans here are always equivalent to running just a WCOJ algorithm.} The \EHP GHDs for TPC-H 
query 5 and matrix multiplication are shown in \Cref{fig:core_components}.

\begin{algorithm}
  \begin{lstlisting}[
        basicstyle = \scriptsize,
        language = C,
        numbers = left,
        mathescape]
  //Input: Hypergraph $H = (V,E)$, and a tuple $t$.
  Generic-Join($V$,$E$,$t$):
    if $|V| = 1$ then return $\cap_{e \in E} R_e[t]$.
    Let $I = \{ v_{1} \}$ // the first attribute.
    Q $\leftarrow \emptyset$ // the return value
    // Intersect all relations that contain $v_{1}$
    // Only those tuples that agree with $t$.
    for every $t_v \in \cap_{e \in E : e \ni v_{1}} \pi_{I} (R_e[t])$ do
      $Q_t$ $\leftarrow$ Generic-Join($V - I$, $E$, $t :: t_v$ )
      Q $\leftarrow Q \cup \{ t_v \} \times Q_t $
    return Q
  \end{lstlisting}
  \caption{Generic Worst-Case Optimal Join Algorithm}
  \label{fig:generic_worst_case}
\end{algorithm}

\subsection{Join Algorithm}
\label{sec:computation_model}

The generic WCOJ algorithm \cite{ngo2014skew} shown in \Cref{fig:generic_worst_case} is the
computational kernel for all queries in the \EHP engine. This algorithm is
used in each node of the GHD-based query plan. The generic WCOJ 
algorithm can be asymptotically better than any pairwise join algorithm \cite{ngo2012worst}. 
Intuitively, the generic WCOJ algorithm joins 
attributes in a multiway fashion as opposed to the traditional approach of
joining relations in a pairwise fashion. This means that
queries in \EHP are processed one attribute equivalence class at a time,
and the core operation to process an equivalence class is a set intersection. 
Selecting the order that these attributes are processed in is
the focus of \Cref{sec:cost_based}. An unrolling of this algorithm on
TPC-H query 5 and matrix multiplication is shown in \Cref{fig:core_components}.

\section{SQL to GHDs}
\label{sec:query_optimization}

The query compilation techniques presented in
\Cref{sec:query_model} are able to capture a wide
range of domains, including linear algebra, message passing, and graph queries
\cite{joglekar2016ajar,aberger2015emptyheaded,aberger2016old}.
However, most of this work has been theoretical and none of the
current literature demonstrates how to capture general SQL-style queries 
in such a framework. In this
section we show how to extend this theoretical work to more complex queries. 
In particular we describe
how \EHP translates generic SQL queries (which \EH could not) to hypergraphs in \Cref
{sec:sql_to_hypergraph}. Using this translation, we show how the 
well-known optimization of attribute elimination is captured both logically and physically in \EHP. 
In \Cref{sec:ghd_optimization} we describe the manner in which \EHP selects a GHD
and optimizes it by pushing down selections.
We argue that with these extensions \EHP represents the first practical 
implementation of these techniques capable of capturing general SQL 
workloads.

\subsection{SQL to Hypergraph Translation}
\label{sec:sql_to_hypergraph}

We demonstrate how certain features of SQL, such as complex
expressions inside aggregation functions, can be translated to operations on annotated
relations using a series of simple rules to construct query hypergraphs.
The rules \EHP uses to translate a 
SQL query to a hypergraph $H = (V, E)$ and an aggregation ordering $\alpha$ are as follows:
\begin{enumerate} \compactify
\item The set $V$ of vertices in the hypergraph contains all of the key columns in
the SQL query. The set of hyperedges $E$ is each relation in the SQL query. \label{item:query_model_vertices} All attributes that appear in an equi-join condition are mapped to the same attribute in $V$. \label{item:query_model_join}
\item All key attributes that do not appear in the output of the query must be in
the aggregation ordering $\alpha$.  \label{item:query_model_agg_order}
\item If only the columns of a single relation appear
inside of an aggregation function, the expression inside of the aggregation function is the annotation of that relation. If none of the relation's columns appear inside an aggregation function, the relation's annotation is the identity element. If the inner expression of an aggregation function touches multiple relations, those relations are constrained to be in the same node of the GHD where the expression is the output annotation. \label{item:query_model_annotation}
\item The rules above do not capture annotations which are not aggregated, so 
these annotations are added to a meta data container $M$ that associates 
them to the hyperedge (relation) from which they originate.
\label{item:query_model_selection}
\end{enumerate}

\begin{example}
Consider how these rules capture TPC-H query 5 from \Cref{fig:core_components}
in a \EHP query hypergraph.

By Rule 1, the equality join in this query is captured in the set of 
vertices ($V$) and hyperedges ($E$) shown in the hypergraph in 
\Cref{fig:core_components}. The columns \verb|c_custkey| and
\verb|o_custkey| must be mapped to the same vertex in $``custkey" \in V$.
 Similarly, the columns
\verb|l_orderkey| and \verb|o_orderkey| are mapped to the vertex $``orderkey"$,
the columns \verb|l_suppkey| and \verb|s_suppkey| are mapped to the vertex
$``suppkey"$, the columns \verb|c_nationkey|, \verb|s_nationkey|, and
\verb|n_nationkey| are mapped to the vertex $``nationkey"$, the columns
\verb|n_regionkey| and \verb|r_regionkey| are mapped to the vertex
$``regionkey"$.

By Rule 2, a valid aggregation ordering is:
\begin{align*}
\alpha = [regionkey, nationkey, suppkey, custkey, orderkey]
\end{align*}
with the aggregation operator $\Sigma$ (the order is irrelevant here).

To apply Rule 3, consider the expression inside the aggregation function, 
\verb|l_extendedprice * (1 - l_discount)|. Only columns on the \verb|lineitem| 
table are involved in this expression, so the annotations on the \verb|lineitem| 
table will be this expression for each tuple. The \verb|orders| and 
\verb|customer| tables do not have any columns in aggregation expressions, so 
they are annotated with the identity element.

By Rule 4, the hypergraph does not capture the 
attributes \verb|n_name|, \verb|o_orderdate|, or \verb|r_name| but our metadata 
container $M$ does. $M$ here is the following: 
\{\verb|n_name| $\leftrightarrow$ \verb|nation|, \verb|r_name| $\leftrightarrow$ \verb|region|, \verb|o_orderdate| $\leftrightarrow$ \verb|orders|\}.

\end{example}

\paragraphafterexample*{Attribute Elimination}
The rules above only add the attributes that are used in the query to the hypergraph.
Although this elimination of unused attributes is obvious, ensuring this physically in the \EHP
trie data structure was non-trivial. To do this we ensured that any number
of levels from the trie can be used during query execution.
This means that annotations can be reached individually from any level of the
trie. Further, the annotations are all stored in individual data buffers 
(like a column store)
to ensure that they can be loaded in isolation. These fundamental differences with
\EH, enable \EHP to support attribute elimination both logically and
physically. This is essential on dense LA kernels because it enables 
\EHP to call BLAS routines by storing a single dense 
annotation in flat (BLAS compatible) buffer.

\subsection{GHD Optimization}
\label{sec:ghd_optimization}

After applying the rules in \Cref{sec:sql_to_hypergraph}, a GHD is selected 
using the process described in \Cref{sec:query_model}. Still, \EHP needs a way 
to select among multiple GHDs that the theory cannot distinguish. In this 
section we explain how \EHP adapts the theoretical definition of GHDs to both
select and produce practical query plans.

\paragraph*{Choosing Among GHDs with the same FHW} 
For many queries, multiple GHDs have the same 
FHW. Therefore, a practical implementation must 
also include principled methods to choose between query plans 
with the same worst-case guarantee. Fortunately, there are three intuitive characteristics 
of  GHD-based query plans that makes this choice relatively simple (and cheap). 
The first is that the smaller a GHD is (in terms of number of nodes and height), the quicker
it can be executed (less generated code). The second is that fewer intermediate
results (shared vertices between nodes) results in faster execution time. Finally,
the lower selection constraints appear in a query plan corresponds to how early
work is eliminated in the query plan. Therefore,
\EHP uses the following order of heuristics to choose between GHDs with the same
FHW:

\begin{enumerate} \compactify
\item Minimize $|V_T|$ (number of nodes in the tree).
\item Minimize the depth (longest path from root to leaf).
\item Minimize the number of shared vertices
 between nodes.
\item Maximize the depth of selections.
\end{enumerate}

Although most of the queries in this paper are single-node
GHDs, on the two node TPC-H query 5, using these rules to select a GHD
results in a \numb{3x} performance advantage over a GHD 
(with the same FHW) that violates the rules above. 
We explain how selections are pushed down below joins in \EHP GHDs next. 

\paragraph*{Pushing Down Selections Below Joins}
Our mechanism of selecting a GHD ensures that selections appear as low as 
possible in a GHD, but we still need a mechanism to push down selections down 
below joins in our chosen GHD. Motivated by how \EH does this 
\cite{aberger2015emptyheaded}, the \EHP query 
compiler pushes down selections below joins by:

\begin{enumerate} \compactify
\item Taking as input an optimal (with respect to FHW) GHD $T = (V_T, E_T)$ 
and a set of selections $\sigma_{a_i}$ applied to attributes $a_i$.
\item For each $\sigma_{a_i}$: Let $e_i$ be the edge that contains the vertex 
derived from $a_i$ or that has the meta data $M$ associated with $a_i$. 
Let $t_i$ be the GHD node in $V_T$ associated with $e_i$. If $t_i$ contains more
than one hyperedge, create a new GHD node $t'_i$ that contains only $e_i$, and
make $t'_i$ a child of $t_i$.
\end{enumerate}

This means that \EHP pushes down selections by creating new GHD
nodes with only the selection constraints under the original GHD
nodes. On TPC-H query 5 this results in the GHD
shown in \Cref{fig:core_components} which executes \numb{1.8x} faster than 
the GHD without this optimization.

\begin{figure*}
 \centering
 \begin{subfigure}{0.30\linewidth}
 \includegraphics[width=\linewidth]{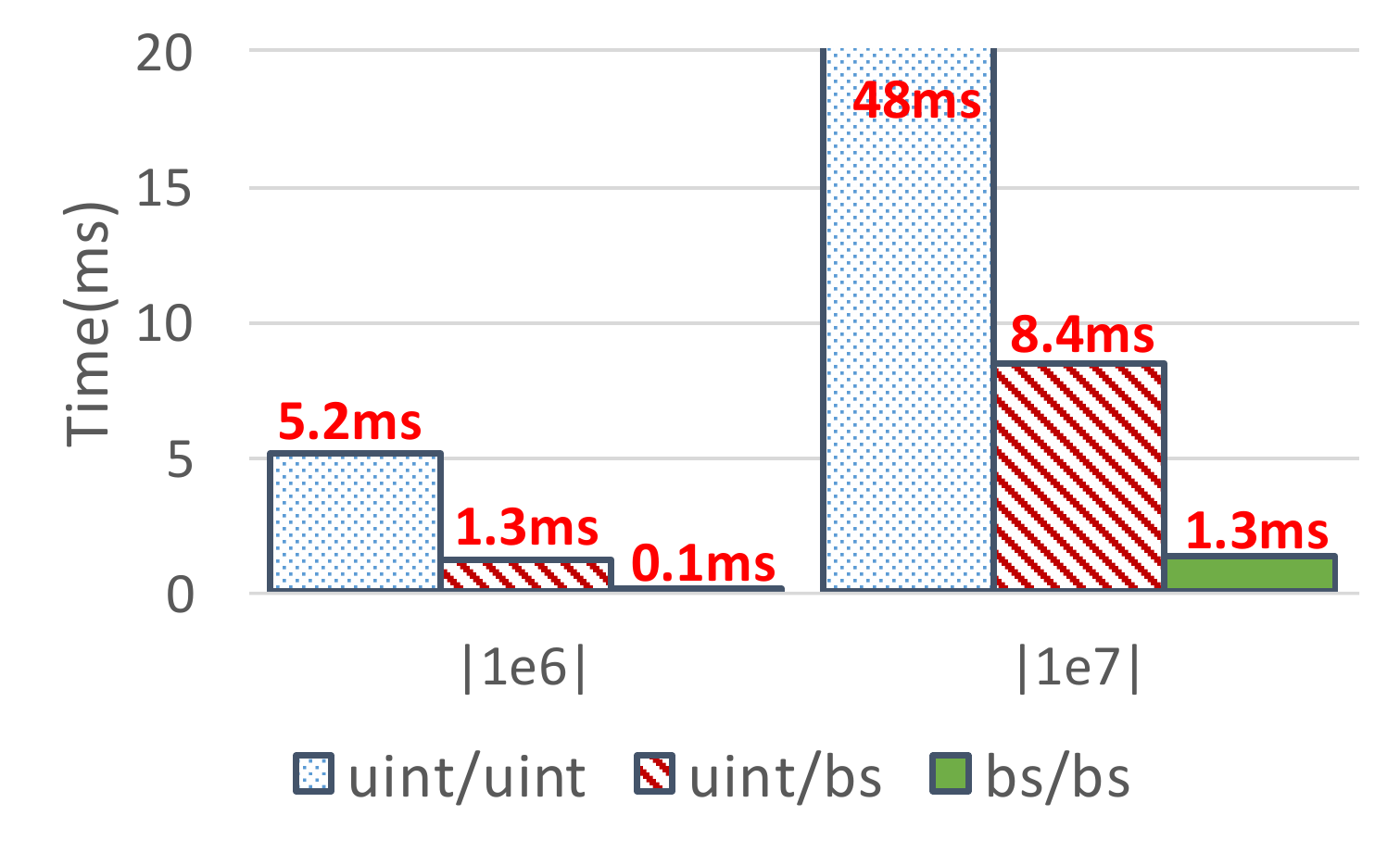}
 \caption{The performance of \texttt{uint} $\cap$ \texttt{uint}, \texttt{uint} $\cap$ \texttt{bs}, and \texttt{bs} $\cap$ \texttt{bs} intersections with cardinalities of 1e6 and 1e7. This is used to derive the intersection cost (\texttt{icost}) estimate for each of type of intersection.}
 \label{fig:intersection_cost}
 \end{subfigure} \hfill
  \begin{subfigure}{0.30\linewidth}
 \includegraphics[width=\linewidth]{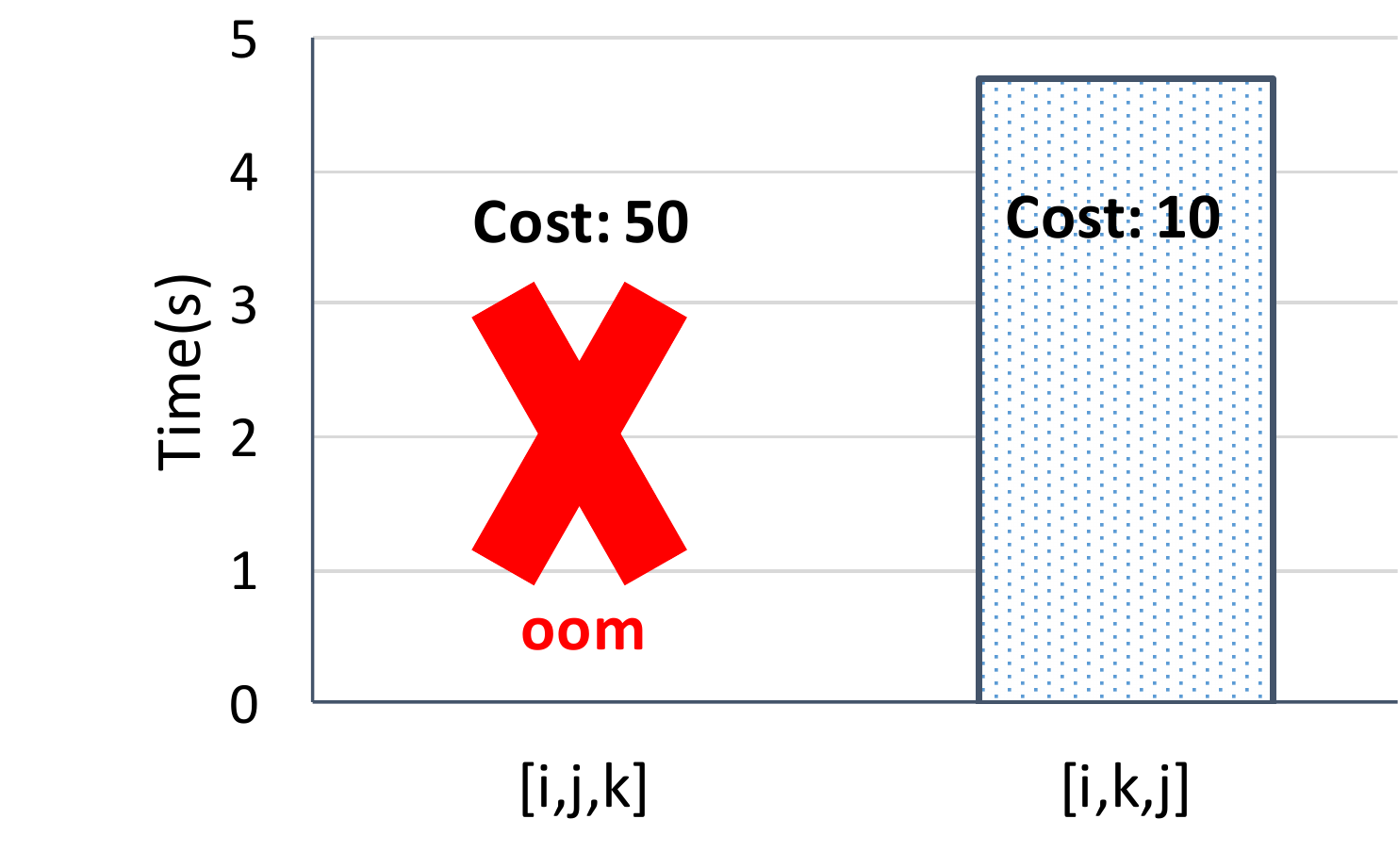}
  \caption{The performance and cost of two attribute orders for sparse matrix multiplication on the nlp240 matrix. The cost 50 order runs out of memory (oom) on a machine with 1TB of RAM.}
  \label{figs:mat_mult_ao}
 \end{subfigure} \hfill
  \begin{subfigure}{0.30\linewidth}
 \includegraphics[width=\linewidth]{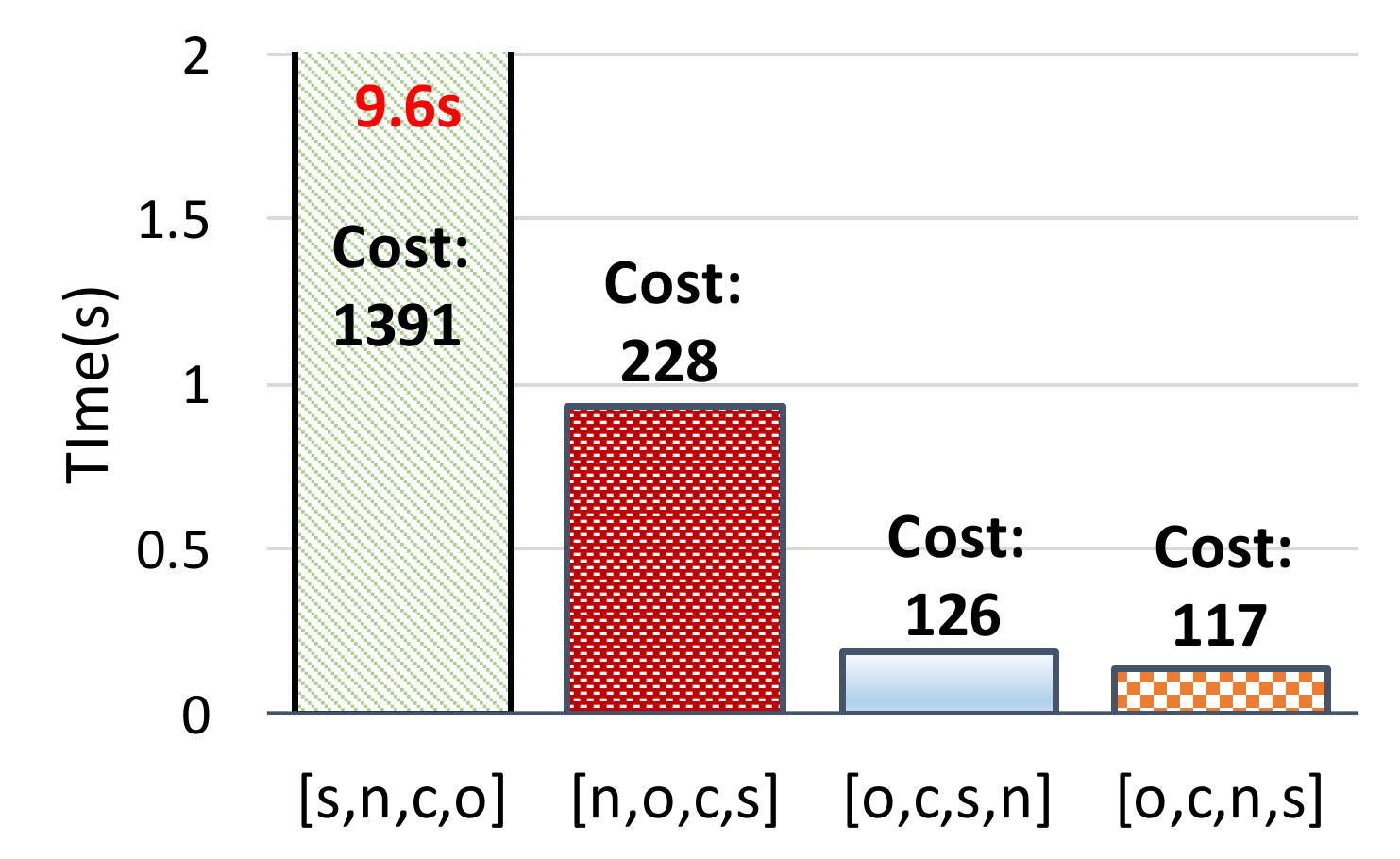}
  \caption{The performance and cost of four attribute orders for the expensive GHD node on TPC-H query 5 at SF 10. Attributes are: o = orderkey, c = custkey, s = nationkey, and n = nationkey.}
    \label{fig:tpch_q5_ao}
 \end{subfigure} \hfill
\caption{Cost estimation experiments.}
\label{fig:intersections}
\end{figure*}

\section{Cost-Based Optimizer}
\label{sec:cost_based}

After a GHD-based query plan is produced using the process described in 
\Cref{sec:query_model,sec:query_optimization}, \EHP needs to select an attribute
order for the WCOJ algorithm. Similar to the classic query optimization problem  
of selecting a join  
order \cite{garcia2008database}, WCOJ attribute ordering can result in 
orders of magnitude performance differences on the same query. Unfortunately, the known 
techniques for estimating the cost of join orders are designed for Selinger-style \cite{astrahan1976system}
query optimizers using pairwise join algorithms---not a GHD-based query optimizer
with a WCOJ algorithm. In this section 
we present the first cost-based optimizer for the generic
WCOJ algorithm and validate the crucial observations it uses to derive its cost 
estimate. We show that this optimizer selects attribute orders 
that provide up to a three orders of magnitude speedup over ones that \EH could 
select. 

\rv{
\paragraph*{Optimizer Overview} \EHP's cost-based optimizer 
selects a key attribute order
for each node in a GHD-based query plan. As is standard \cite{aberger2015emptyheaded}, 
\EHP requires that 
materialized key attributes appear before those that are projected away (with one 
important exception described in \Cref{sec:relaxing}) and that materialized attributes always adhere to some global ordering (e.g. if attribute `a' is before attribute `b' in one GHD node order, then 'a' must be before 'b' in all GHD orders). To assign an attribute order to each GHD node, \EHP's cost-based optimizer: traverses the GHD in a top-down fashion,
considers all attribute orders\footnote{\small We
remind the reader that the number of attributes considered here is only the number 
of key (joined) attributes inside of a GHD node which is typically small.} adhering to the previously described criteria at each node, and selects the attribute order with the lowest cost estimate.
}
For each order, a cost estimate is computed 
based on two characteristics of the generic WCOJ algorithm: (1) the algorithm 
processes one attribute at a time and (2) set intersection is the bottleneck 
operator. As such, \EHP assigns a set intersection cost (\texttt{icost}) and a cardinality
weight (\Cref{sec:weights}) to each key attribute (or vertex) in the 
hypergraph of a GHD node. Using this, the cost estimate for a given a key attribute (or hypergraph vertex) order 
[$v_0$,$v_1$,...,$v_{|V|}$] is:
$$ \mathrm{cost} = \sum_{i=0}^{|V|} \left(\mathrm{\texttt{icost}}(v_i) \times \mathrm{weight}(v_i)\right) $$
The remainder of this section discusses how the \texttt{icost}s 
(\Cref{sec:intersection_cost}) and weights (\Cref{sec:weights}) are derived.

\subsection{Intersection Cost}
\label{sec:intersection_cost}
The bottleneck of the generic WCOJ algorithm is set intersection operations. 
In this section, we describe how to derive a simple cost estimate, called 
\texttt{icost}, for the set intersections in the generic WCOJ algorithm.

\subsubsection{Cost Metric}

Recall that the sets in \EHP tries are stored using an unsigned integer 
layout (\texttt{uint}) if they are sparse and a bitset layout 
(\texttt{bs}) if they are dense, a design inherited from \EH. Thus, the 
intersection algorithm used is different depending on the data layout of the 
sets. These different 
layouts have a large impact on the set intersection performance,
even with similar cardinality sets. For example, 
\Cref{fig:intersection_cost}, shows that a \texttt{bs} $\cap$ \texttt{bs} is roughly 
50x faster than a \texttt{uint} $\cap$ \texttt{uint} with the same cardinality sets. 
Therefore, \EHP uses the results from 
\Cref{fig:intersection_cost} to assign the following 
\texttt{icost}s:
\begin{center}
\texttt{icost}(\texttt{bs} $\cap$ \texttt{bs}) = 1, 
\texttt{icost}(\texttt{bs} $\cap$ \texttt{uint}) = 10, \\
\texttt{icost}(\texttt{uint} $\cap$ \texttt{uint}) = 50 \\
\end{center}

Unfortunately, it is too expensive for the query compiler to check (or track)
the layout of each set during query compilation---set layouts 
are chosen dynamically during data ingestion and a single trie can have millions
of sets. To address this \EHP uses \Cref{principle:trie_levels}.

\begin{principle}
\label{principle:trie_levels}
The sets in the first level of a trie are typically dense and therefore
represented as a bitset. The sets of any other level of a trie are typically 
sparse (unless the relation is completely dense) and therefore represented 
using the unsigned integer layout.

\vspace{1.5mm}
\noindent \textbf{Empirical Validation:} Consider a 
trie for the TPC-H \texttt{lineitem} relation where the trie levels
correspond to the key attributes 
[\texttt{l\_orderkey},\texttt{l\_suppkey},\texttt{l\_partkey},\texttt{l\_linenumber}] 
in that order. At scale factor 10, each level of this trie has the following number
of \texttt{uint} and \texttt{bs} sets:
\end{principle}  

{
\small
\begin{itemize}
    \item 1st level(\texttt{l\_orderkey}) = \{0 \texttt{uint} sets, 1 \texttt{bs} set\}
    \item 2nd level(\texttt{l\_suppkey}) = \{14999914 \texttt{uint} sets, 86 \texttt{bs} sets\}
    \item 3rd level(\texttt{l\_partkey}) = \{59984817 \texttt{uint} sets, 0 \texttt{bs} sets\}
    \item 4th level(\texttt{l\_linenumber}) = \{59986042 \texttt{uint} sets, 0 \texttt{bs} sets\}
\end{itemize}  
}

Thus, given a key attribute order [$v_0$,...,$v_{|V|}$] (where each 
$v_i \in V)$, the \EHP optimizer assigns an \texttt{icost} 
to each $v_i$, in order of appearance, using the following method which leverages
\Cref{principle:trie_levels}:

\begin{itemize} \compactify
	\item For each edge $e_j$ with node $v_i$, assign $l(e_j)$ (where 
    $l$=layout), to either \texttt{uint} or \texttt{bs}. 
    As a reminder, edges are relations and
    vertices are attributes. Thus, for each relation this assignment guesses one 
    data layout for all of the relation's $v_i$ sets. 
    If $e_j$ has been assigned with a previous vertex $v_{k}$ where k $<$ i, 
    $l(e_j) = \mathrm{\texttt{uint}}$
    (not the first trie level), 
    otherwise 
    $l(e_j) = \mathrm{\texttt{bs}}$. 
	\item Compute the cost of intersecting the $v_i$ attribute from each edge (relation) $e_j$.
	For a vertex with two edges, the pairwise \texttt{icost} is used. 
    For a vertex with $N$ edges, where 
	$N > 2$, the \texttt{icost} is the sum of pairwise \texttt{icost}s where 
    the \texttt{bs} sets are \emph{always} processed first. For example,
    when $N=3$ and l($e_0$) $\leq$ l($e_1$) $\leq$ l($e_2$) where \texttt{bs} $<$ \texttt{uint}, $\mathrm{\texttt{icost}} =  \mathrm{\texttt{icost}}( \mathrm{l}(e_0) \cap  \mathrm{l}(e_1)) +  \mathrm{\texttt{icost}}(\mathrm{l}(\mathrm{l}(e_0) \cap \mathrm{l}(e_1)) \cap \mathrm{l}(e_2))$. Note, 
    $\mathrm{\texttt{uint}} = \mathrm{l}(\texttt{bs} \; \cap \; \texttt{uint})$. 
\end{itemize}

\begin{example}
Consider the attribute order [\texttt{orderkey},\texttt{custkey},\texttt{nationkey},\texttt{suppkey}] for 
one of the GHD nodes in TPC-H query 5 (see \Cref{fig:core_components}). 
The \texttt{orderkey} vertex is assigned an \texttt{icost} of 1 as it is 
classified with \texttt{[bs $\cap$ bs]} intersections. The \texttt{custkey}
vertex is assigned an \texttt{icost} of 10, classified with 
\texttt{[uint $\cap$ uint]} intersections. The \texttt{nationkey} 
vertex is assigned an \texttt{icost} of 11, classified with 
\texttt{[bs $\cap$ bs $\cap$ uint]} intersections. Finally, 
the \texttt{suppkey} vertex is assigned an 
\texttt{icost} of 50, classified with \texttt{[uint $\cap$ uint]} 
intersections. 
\end{example}

Finally, in the special case of a completely dense relation, the LevelHeaded 
optimizer assigns an \texttt{icost} of 0 because no set intersection is necessary in this 
case. This is essential to estimate the cost of LA queries properly.

\subsubsection{Relaxing the Materialized Attributes First Rule}
\label{sec:relaxing}
An interesting aspect of the intersection cost metric is that the cheapest key 
attribute order could have materialized key attributes come after 
those which are projected away.\footnote{\small We remind the reader that later
(or after) in the attribute order corresponds to a lower level of the tries.} To
support such key attribute orders, the execution engine must be able to combine children 
(in the trie) of projected away key attributes using a set union or \texttt{GROUP BY} 
(to materialize the result sets). Unfortunately, it is
difficult to design an efficient data structure to union more than one
level of a trie (materialized key attribute) in parallel (e.g. use a massive 2-dimensional
buffer or incur expensive merging costs). Therefore, \EH kept its design simple 
and never considered relaxing the
rule that materialized attributes must appear before those which are
projected away. In \EHP we relax this rule by allowing
1-attribute unions (see \Cref{sec:query_execution}) on keys when it can lower 
the \texttt{icost}.

Within a GHD node, \EHP relaxes the materialized attributes first rule under 
the following conditions:
\begin{enumerate} \compactify
\item The last attribute is projected away.
\item The second to last attribute is materialized.
\item The \texttt{icost} is improved by swapping the two attributes. 
\end{enumerate}
These conditions ensure that 1-attribute union will only be introduced when
the \texttt{icost} can be lowered. 

\begin{figure*}
 \centering
 \begin{subfigure}{0.30\linewidth}
 \includegraphics[width=\linewidth]{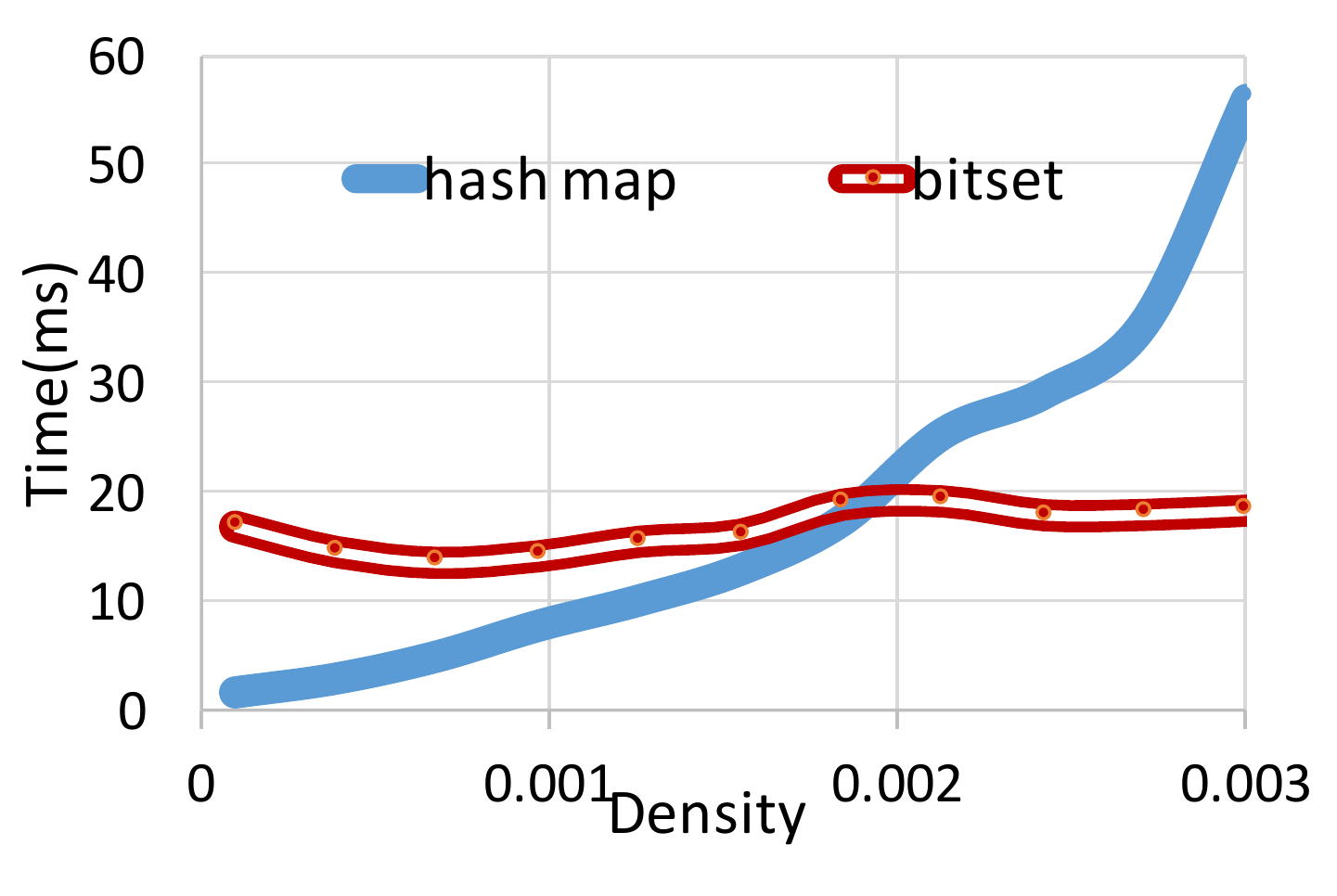}
   \caption{The performance of \EHP's hash map and bitset \texttt{GROUP BY} operators for key 
   values for various output set densities. The range of output values is fixed to [0,1e6].}
   \label{fig:group_by_keys}
 \end{subfigure} \hfill
  \begin{subfigure}{0.30\linewidth}
 \includegraphics[width=\linewidth]{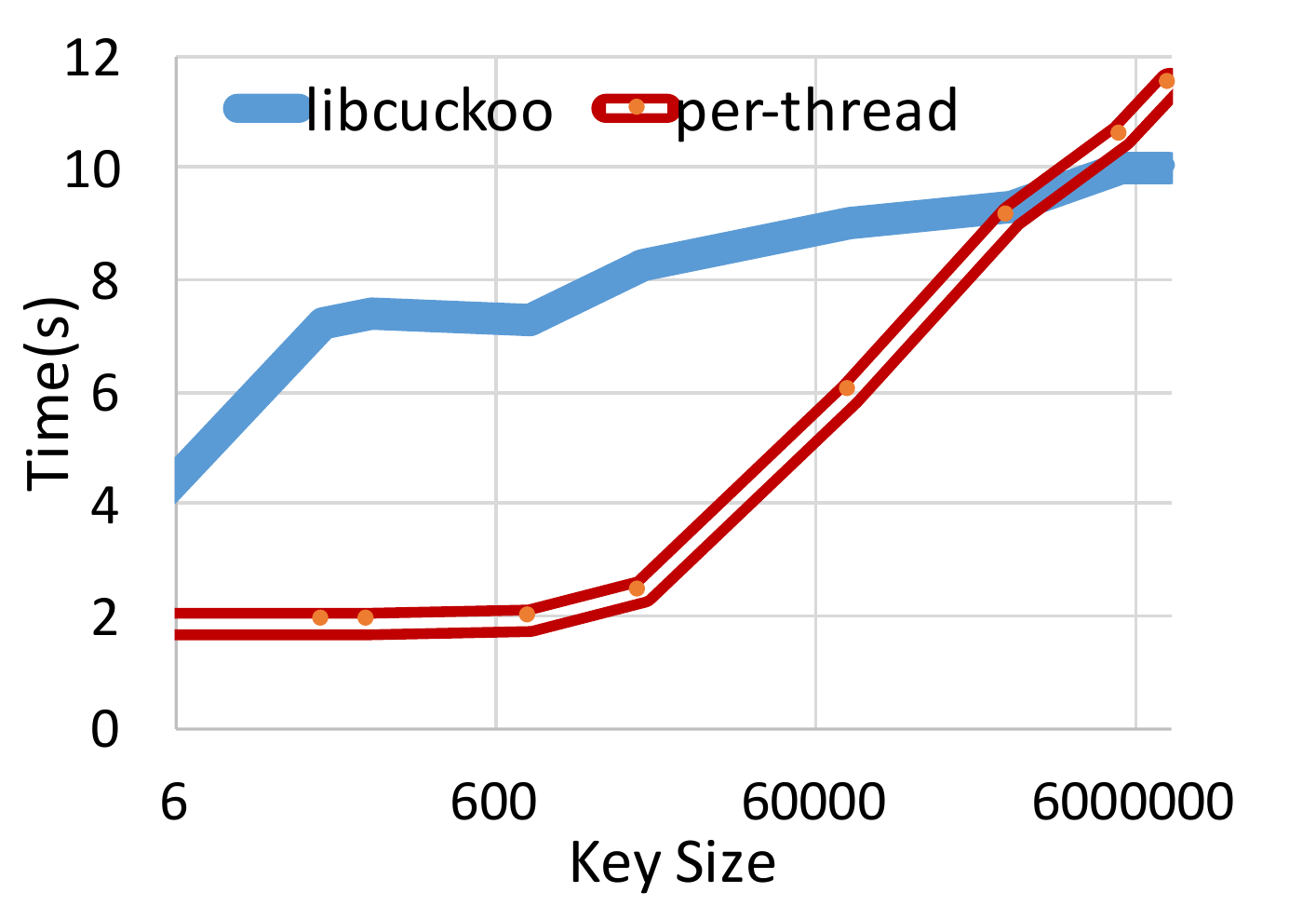}
    \caption{The performance of 1e7 parallel upsert operations for libcuckoo and a collection of per-thread hash maps with different output hash map key sizes. The \texttt{GROUP BY} is on a single \texttt{uint} attribute. 56-threads are used.
     }
     \label{fig:group_by_anno_simple}
 \end{subfigure} \hfill
  \begin{subfigure}{0.30\linewidth}
 \includegraphics[width=\linewidth]{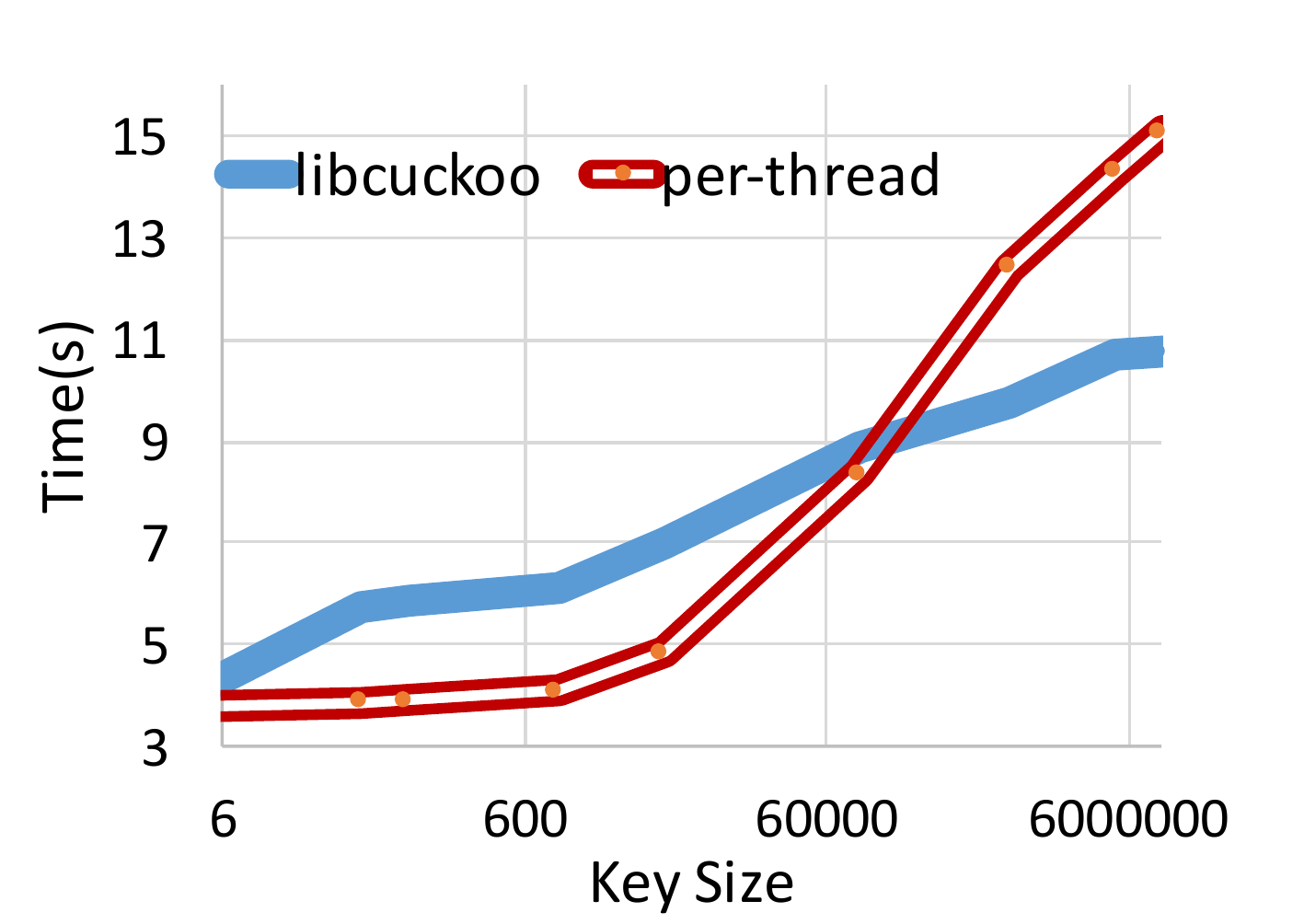}
     \caption{The performance of 1e7 parallel upsert operations for libcuckoo and a collection of per-thread hash maps with different output hash map key sizes. The \texttt{GROUP BY} is on 2 \texttt{uint} and 4 \texttt{double} attributes. 56-threads are used.}
     \label{fig:group_by_anno_complex}
 \end{subfigure} \hfill
\caption{The performance of \EHP's \texttt{GROUP BY} implementations on key and annotation attributes.}
\vspace{-1mm}
\label{fig:unions}
\end{figure*}

\begin{example}
 Consider the sparse matrix multiplication query and its unrolling
 of the generic WCOJ algorithm  for an attribute order of 
 \texttt{[i,j,k]} shown in \Cref{fig:core_components}. This attribute order has a
 cost 50 \texttt{[uint $\cap$ uint]} assigned to the \texttt{k} attribute.

Now consider an attribute
order  \texttt{[i,k,j]}. Here a cost 10 \texttt{[uint $\cap$ bs]} is assigned to \texttt{k} and the 
unrolling of the generic WCOJ algorithm (where () denotes accesses to keys and [] denotes access 
to annotations) is the following:
{
\small
\begin{tabbing}
for \= $i \in \pi_{i}M1$ do\\
\>$s_j \leftarrow \emptyset$ \\
\>for \= $k \in \pi_{k}M1(i) \cap \pi_{k}M2$ do\\
\>\>for \= $j \in \pi_{j}M2(k)$ do\\
\>\>\>$s_j[j] \; += \; \pi_{v1}M1[i,k] * \pi_{v2}M2[k,j]$ \\
\> $out(i) \leftarrow s_j$
\end{tabbing}
}
This lower cost attribute order 
recovers the same loop ordering as Intel MKL on sparse matrix multiplication 
\cite{patwary2015parallel} and its bottleneck is the 
\texttt{+=} operation that unions \texttt{j} values and sums  \texttt{v1*v2} values. 
This is in contrast to the standard bottleneck operation of a set intersection as in the 
\texttt{[i,j,k]} order. In \Cref{sec:query_execution}
we discuss the tradeoffs around implementing this union add or \texttt{GROUP BY} 
operation. \Cref{figs:mat_mult_ao} shows that this order is essential to run 
sparse matrix multiplication as a join query without running out of memory.
\end{example}

\subsection{Weights}
\label{sec:weights}

Like classic query optimizers, \EHP also tracks the cardinality of 
each relation as this influences the \texttt{icost}s
for the generic WCOJ algorithm. \Cref{fig:intersections} shows the 
unsurprising fact that larger cardinality sets result in longer intersection 
times. To take this into account when computing a cost estimate \EHP
assigns weights to each vertex using \Cref{principle:cardinalty}, which directly
contradicts conventional wisdom from pairwise optimizers:

\begin{principle}
\label{principle:cardinalty}
The highest cardinality attributes should be processed first in the generic 
WCOJ algorithm. This enables these attributes to partake in
fewer intersections (outermost loops) and ensures that they are a higher 
trie levels (more likely \texttt{bs}'s with lower 
\texttt{icost}s). 

\vspace{1.5mm}
\noindent \textbf{Empirical Validation:} We show in \Cref{fig:tpch_q5_ao} 
that the generic WCOJ can run over \numb{70x} faster on TPC-H query 5 when 
the high cardinality \texttt{orderkey} attribute 
is first in the attribute order instead of last.
\end{principle}

\EHP's 
goal when assigning weights is to follow \Cref{principle:cardinalty} by 
assigning high cardinality attributes heavier weights so that they appear 
earlier in the attribute order. To do this, \EHP assigns a cardinality score
to each queried relation and uses this to weight to each attribute. We describe
this in more detail next.

\paragraph*{Score} \EHP maintains a cardinality score for each 
relation in a query which is just the relation's cardinality relative to the 
highest cardinality relation in the query. The score (out of 100) for 
a relation $r_i$ is: 
$$ \mathrm{score} = \mathrm{ceiling}\left(\frac{|r_i|}{|r_{heavy}|}\times100\right)$$
where $r_{heavy}$ is the highest cardinality relation in the query.
\paragraph*{Weight} To assign a weight to each vertex \EHP uses the highest score edge (or 
relation) with the vertex when a high selectivity (equality) constraint is present, otherwise \EHP takes the lowest score edge (or relation).  
The intuition for 
using the highest score edge (or relation) with a high 
selectivity constraint is that this relation represents the amount of work that
could be filtered (or eliminated) at this vertex (or attribute).
The intuition for otherwise taking the lowest score edge (or relation) is that the 
output cardinality of an intersection is at most the size of the smallest set. 
\begin{example}
Consider TPC-H Q5 at scale factor 10. The cardinality score for each relation 
here is:
\begin{center}
\small
score(lineitem) = 100, score(orders) = 26, score(customer)= 3, \\ 
score(region) = 1, score(supplier) = 1, score(nation) = 1 \\
\end{center}
The weight for each vertex is (region is equality selected):
\begin{center}
\small
weight(orderkey) = min(26,100), weight(custkey) = min(3,26)  \\
weight(suppkey) = min(1,100), weight(nationkey) = min(1,1,3) \\
weight(regionkey) = max(1,1) \\
\end{center}
These weights are then used to derive the cost estimates shown in \Cref{fig:tpch_q5_ao}.
\end{example}

\section{Group By Tradeoffs}
\label{sec:query_execution}

Designing a skew-resistant \texttt{GROUP BY} operator requires careful 
consideration of the tradeoffs associated with its implementation. In this section we present these tradeoffs in the context
of \EHP. Although many \texttt{GROUP BY}s are automatically captured in \EHP's 
tries, two classes of \texttt{GROUP BY}s (that were not supported in \EH) require 
an additional implementation in \EHP: 
 (1) a \texttt{GROUP BY} to union 
keys in the attribute orders from \Cref{sec:relaxing} and (2) a \texttt{GROUP BY} 
on annotations. In this section we present the tradeoffs around two implementations 
for each type of \texttt{GROUP BY} and describe a simple optimizer that automatically exploits these tradeoffs to 
select the implementation used during execution. In \Cref{sec:experiments}, we show that this \texttt{GROUP BY} optimizer 
provides up to a \numb{875x} and \numb{185x} speedup over using a single \texttt{GROUP BY} implementation
on BI and LA queries respectively. 

\paragraph*{GROUP BY Key} An important artifact of the key attribute orders from
\Cref{sec:cost_based} is that a projected away attribute can appear 
before a materialized attribute in \EHP. In this case a \texttt{GROUP BY} 
on a key attribute is needed to union the result. 
 \EHP selects between two implementations for this: (1) a hash map that 
 upserts (key, value) pairs as they are encountered and (2) a bitset 
 that unions (OR's) keys and a dense array to hold the values.
\Cref{fig:group_by_keys} shows that the performance of these implementations 
depends greatly on the density of the 
output key attribute set. The bitset performs better when the output density is high and the
hash map performs better when the output density is low. To predict the output 
set's density \EHP leverages a simple
observation: the output density is correlated in a 
1-1 manner with the density of 
projected away attribute (set being looped over). As such, \EHP uses the
density of the projected away attribute to predict the 
output set's density and select the implementation used.

\paragraph*{GROUP BY Annotation} \EHP supports SQL queries with \texttt{GROUP BY}
operations on annotations or on both annotations and keys.
To support this in parallel, \EHP selects between two implementations of a 
parallel hash map: (1) libcuckoo \cite{li2014algorithmic} and (2) a per-thread
instance of the C++ standard library unordered map. In the context of a \texttt{GROUP BY},
the crucial operation for these hash maps is an upsert and we found that each implementation
worked best under different conditions. \Cref{fig:group_by_anno_simple,fig:group_by_anno_complex}
show that when the output key size is small (many collisions), the per-thread 
implementation  
outperforms libcuckoo by up to 4x. In contrast, libuckoo outperforms 
the per-thread approach by an order of magnitude when the output size
is large (few collisions).  Unfortunately, predicting output cardinalities is 
a difficult problem \cite{schiefer1998method}, so \EHP leverages a general 
observation to avoid this when selecting between these two approaches: 
libcuckoo is at worst close ($<$2x) to the 
performance of the 
per-thread approach when the hash-map key tuple is wide ($>$3 values), and the 
per-thread  approach is at worst close ($<$2x) to the performance of libcuckoo 
when the hash map key is small ($<$3 values). This trend is shown in 
\Cref{fig:group_by_anno_simple,fig:group_by_anno_complex}. Thus, the \EHP optimizer 
selects
a per-thread hash map when the hash map key is a tuple of three elements or less 
and 
libcuckoo otherwise.

\vspace{-1.4mm}
\section{Experiments}
\label{sec:experiments}

We compare \EHP to state-of-the-art relational database management 
engines and LA packages on standard BI and LA benchmark queries. 
We show that \EHP is able to compete within \numb{2.5x} of these 
engines, while sometimes outperforming them, and that the techniques from 
in \Cref{sec:query_optimization,sec:cost_based,sec:query_execution} can provide
up to a three orders of magnitude speedup. This validates that 
a WCOJ architecture is a practical solution for both BI and LA queries.

\subsection{Setup}
We describe the experimental setup for all experiments.
\vspace{-1mm}
\paragraph*{Environment} 
\EHP is a shared memory engine that runs and is evaluated on a single node server. 
As such, we ran all experiments on a single machine with a total of 56 cores 
on four Intel Xeon E7-4850 v3 CPUs and 1 TB of RAM. For all engines, we chose 
buffer and heap sizes that were at least an order of magnitude larger than the 
dataset to avoid garbage collection and swapping data out to disk.

\begin{table*}[!ht]
\scriptsize
  \begin{center}
    \setlength{\tabcolsep}{9pt}
    \begin{tabular}{@{}llll|rrrrr@{}}
    \toprule
     & Query & Data &Baseline  &\EHP &Intel MKL &HyPer &MonetDB &LogicBlox \\
    \midrule
    \multirow{21}{*}{TPC-H}
    & \multirow{3}{*}{Q1}
        &SF 1    &12ms  &1.79x			&-    &\textbf{1x} &30.59x &74.17x \\
    &   &SF 10   &84ms  &1.73x			&-    &\textbf{1x} &17.86x &23.45x \\
    &   &SF100   &608ms &1.78x			&-    &\textbf{1x} &80.43x &26.12x \\
    \cline{2-9}
    & \multirow{3}{*}{Q3}
        &SF 1    &29ms  &1.11x			&-    &\textbf{1x} &5.56x  &48.28x \\
    &   &SF 10   &111ms &\textbf{1x}	&-    &1.45x &9.88x  &32.59x \\
    &   &SF100   &963ms &1.01x			&-    &\textbf{1x} &9.76x  &10.99x \\
    \cline{2-9}
    & \multirow{3}{*}{Q5}
        &SF 1    &19ms  &1.49x	  		&-    &\textbf{1x} &6.54x  &109x \\
    &   &SF 10   &92ms  &1.40x	  		&-    &\textbf{1x} &4.84x  &55.33x \\
    &   &SF100   &867ms &1.21x	  		&-    &\textbf{1x} &4.04x  &21.33x \\
    \cline{2-9}
    & \multirow{3}{*}{Q6}
        &SF 1    &5ms   &1.73x 			&-    &\textbf{1x} &12.27x &270x \\
    &   &SF 10   &34ms  &1.50x 			&-    &\textbf{1x} &6.65x  &101x \\
    &   &SF100   &283ms &1.61x 			&-    &\textbf{1x} &7.42x  &73.43x \\
    \cline{2-9}
    & \multirow{3}{*}{Q8}
        &SF 1    &16ms   &\textbf{1x} &-    &2.78x &7.96x  &72.77x \\
    &   &SF 10   &45ms   &1.74x 		&-    &\textbf{1x} &15.16x &73.78x \\
    &   &SF100   &1.06ms &1.88x 		&-    &\textbf{1x} &21.55x &25.02x \\
    \cline{2-9}
    & \multirow{3}{*}{Q9}
        &SF 1    &27ms   &\textbf{1x} &-    &1.84x 	   &4.23x  &97.62x \\
    &   &SF 10   &115ms  &\textbf{1x} &-    &4.05x       &4.14x  &57.84x \\
    &   &SF100   &1020ms &\textbf{1x} &-    &5.71x       &5.19x  &21.78x \\
    \cline{2-9}
    & \multirow{3}{*}{Q10}
        &SF 1    &32ms  &1.36x 			&-    &\textbf{1x} &5.88x  &31.56x \\
    &   &SF 10   &196ms &1.26x 			&-    &\textbf{1x} &6.12x  &18.06x \\
    &   &SF100   &869ms &1.78x 			&-    &\textbf{1x} &9.9x   &7.79x \\
    \hline
    \multirow{12}{*}{Linear Algebra}
    & \multirow{3}{*}{SMV}
        &Harbor &2.66ms   	&\textbf{1x} 	&2.89x 		 &10.81x 	&30.80x 	&89.74x \\
    &  	&HV15R 	&68.01ms  	&2.43x 			&\textbf{1x} &25.82x  	&26.47x		&40.72x \\
    &  	&NLP240 &114.97ms 	&1.49x 			&\textbf{1x} &17.23x 	&53.93x 	&113x \\
    \cline{2-9}
    & \multirow{3}{*}{SMM}
        &Harbor &110ms   	&1.63x 			&\textbf{1x} &13.10x 	&27.27x 	&112x\\
    &  	&HV15R 	&18.79s  	&1.35x 		&\textbf{1x} &oom  		&t/o			&48.11x\\
    &  	&NLP240 &1.92s   	&2.44x 			&\textbf{1x} &4.91x 	&t/o 			&78.70x\\
    \cline{2-9}
    & \multirow{3}{*}{DMV}
        &8192 	&7.96ms		&\textbf{1x}   	&\textbf{1x} &4.34x 	&55.14x 	&121x \\
    &  	&12288 	&13.5ms 	&\textbf{1x}   	&\textbf{1x} &5.78x 	&88.89x 	&330x \\
    &  	&16384 	&23.45ms 	&\textbf{1x}   	&\textbf{1x} &18.13x 	&51.18x 	&587x \\
    \cline{2-9}
    & \multirow{3}{*}{DMM}
        &8192 	&2.76s 		&1.02x   		&\textbf{1x} &oom 		&t/o 			&t/o \\
    &  	&12288 	&4.43s  	&1.01x   		&\textbf{1x} &oom 		&t/o  			&t/o\\
    &  	&16384 	&9.29s 	&1.01x   		&\textbf{1x} &oom 		&t/o 			&t/o \\
    \bottomrule
    \end{tabular}
    \caption{Runtime for the best performing engine (``Baseline'') and relative runtime for comparison engines. `-' indicates that the engine did not provide support for the query. `t/o' indicates the system timed out and ran for over 30 minutes. `oom' indicates the system ran out of memory. \label{table:numbers}}
  \end{center}
  \vspace{-5mm}
\end{table*}

\paragraph*{Relational Comparisons} We compare to HyPer, MonetDB, and LogicBlox 
on all queries to highlight the performance of other relational 
databases. Unlike \EHP, these engines are unable to compete within an 
order of magnitude of the best approaches on BI and LA queries. We compare to 
HyPer v0.5.0 \cite{kemper2011hyper} as HyPer is a state-of-the-art in-memory 
RDBMS design. We also compare to the MonetDB Dec2016-SP5 release. MonetDB
is a popular columnar store database engine and is a widely used 
baseline \cite{idreos2012monetdb}. Finally, we compare to LogicBlox v4.4.5 as 
LogicBlox is 
the first general purpose commercial engine to provide similar worst-case 
optimal join guarantees \cite{aref2015design}. Our setup of LogicBlox  
was aided by a LogicBlox engineer. Because of this, we know that LogicBlox 
does not use a WCOJ algorithm for many of the join queries in the 
TPC-H benchmark.
HyPer, MonetDB, and LogicBlox are 
full-featured commercial strength systems (support transactions, etc.) and 
therefore incur inefficiencies that \EHP does not.

\paragraph*{Linear Algebra Package Comparison} We use Intel MKL v121.3.0.109 as the 
specialized linear algebra baseline. This is the best baseline for LA performance 
on Intel CPUs (as we use in this paper). Others \cite{smith2014anatomy} have shown 
that it takes considerable effort and tedious low-level optimizations to approach 
the performance of such libraries on LA queries.

\paragraph*{Omitted Comparisons} We omit a comparison to array databases like SciDB 
\cite{brown2010overview} as these engines call BLAS or LAPACK libraries (like Intel MKL) on the
LA queries that we present in this paper. Therefore, Intel MKL was the 
proper baseline for our LA comparisons. Additionally, SciDB is not 
designed to process TPC-H queries.

\paragraph*{Metrics}  For end-to-end performance, we measure the wall-clock time 
for each system to complete each query. We repeat each measurement seven times, 
eliminate the lowest and the highest value, and report the average. This 
measurement excludes the time used for outputting the result, data statistics 
collection, and index creation for all engines. We omit the data loading and 
query compilation for all systems except for LogicBlox and MonetDB where this 
is not possible. The query compilation time for these queries, especially at large
scale factors, is negligible to the execution time. Still, we found that \EHP's 
unoptimized query compilation process performed within \numb{2x} of HyPer's.
To minimize unavoidable differences with disk-based engines 
(LogicBlox and MonetDB) we place each database in the \emph{tmpfs} in-memory 
file system and collect hot runs back-to-back. Between measurements for the 
in-memory engines (HyPer and Intel MKL), we wipe the caches and 
re-load the data to avoid the use  of intermediate results.

\subsection{Experimental Results}

We show that \EHP can compete within \numb{2x} of HyPer on seven TPC-H queries and 
within \numb{2.5x} of Intel MKL on four LA queries, while outperforming
MonetDB and LogicBlox by up to two orders of magnitude.

\subsubsection{Business Intelligence}

On seven queries from the TPC-H benchmark we show that \EHP can compete 
within \numb{2x} of HyPer while outperforming 
MonetDB by up to \numb{an order of magnitude} and LogicBlox by up to \numb{two orders 
of magnitude}.

\paragraph*{Datasets} We run the TPC-H queries at scale factors 1, 10, 
and 100. We stopped at TPC-H 100 as in-memory engines, such as HyPer, 
often use 2-3x more memory than the size of the input database during loading---therefore 
approaching the memory limit of our machine. For reference, on TPC-H query 1 HyPer uses
161GB of memory whereas \EHP uses 25GB (only loading the data it needs from disk).

\paragraph*{Queries} We choose TPC-H queries 1, 3, 5, 6, 8, 9 and 10 to 
benchmark, as these queries exercise the core operations 
of BI querying and also containing interesting join 
patterns (except 1 and 6). TPC-H queries 1 and 6 do not contain a join and 
demonstrate that although \EHP is designed for join queries, it can also 
compete on scan queries. 

\paragraph*{Discussion} In \Cref{table:numbers} we show that \EHP can outperform 
MonetDB by up to \numb{80x} and LogicBlox by up to \numb{270x} while remaining 
within \numb{1.88x} of the highly optimized HyPer database. Unsurprisingly, the 
queries where \EHP is the farthest off the performance of the HyPer engine is 
TPC-H queries 1 and 8. Here the output cardinality is small and the runtime
is dominated by the \texttt{GROUP BY} operation.
As described in \Cref{sec:query_execution}, \EHP leverages existing hash map 
implementations 
to perform these \texttt{GROUP BY}s, 
and a custom built (NUMA aware) hash map would likely close this performance gap. 
On queries 3 and 9, where the output cardinality is larger (and closer to 
worst-case),  \EHP is able to compete within 11\% of Hyper and sometimes 
outperforms it. In \EHP, we also tested additional optimizations
like indexing annotations with selection
constraints and pipelining intermediate results between GHD nodes. We found that 
these optimizations could 
provide up to a 2x performance increase on queries like TPC-H query 8, but
did not consistently outweigh their added complexity.

\subsubsection{Linear Algebra}

We show that \EHP can compete within \numb{2.5x} of Intel MKL while outperforming 
HyPer and LogicBlox more than \numb{18x} and \numb{587x} respectively on 
LA benchmarks.

\paragraph*{Datasets} We evaluate linear algebra queries on three dense 
matrices and three sparse matrices. The first sparse matrix dataset we 
use is the Harbor dataset, which is a 3D CFD model of the Charleston Harbor
\cite{davis2011university}. The Harbor dataset is a sparse matrix 
\cite{patwary2015parallel} that contains 46,835 rows and columns and 
2,329,092 nonzeros. The second sparse matrix dataset we use is the 
HV15R dataset, which is a CFD matrix of a 3D engine fan \cite{davis2011university}.
The HV15R matrix contains 2,017,169 rows and columns and 283,073,458 nonzeros 
and is a large, non-graph, sparse matrix. 
The final sparse matrix dataset we use is the nlpkkt240 dataset with 27,993,600 
rows and columns and 401,232,976 nonzeros \cite{schenk2008inertia}. The 
nlpkkt240 dataset is a symmetric indefinite KKT matrix.
For dense matrices, we use synthetic matrices with dimensions 
of 8192x8192 (8192), 12288x12288 (12288), and 16384x16384 (16384).

\paragraph*{Queries} We run matrix dense vector multiplication and matrix 
multiplication queries on both sparse (SMV,SMM) and dense (DMV,DMM) matrices. 
These queries were chosen because they are simple to express using joins 
and aggregations in SQL and are the core operations for 
most machine learning algorithms. Further, 
Intel MKL is specifically designed to process these queries and, as a result, 
achieves the largest speedups over using a RDBMS here. 
Thus, these queries represent the most 
challenging LA baseline queries possible.  For both SMM and DMM we
 multiply the matrix by itself, as is standard for 
benchmarking \cite{patwary2015parallel}.

\paragraph*{Discussion} \Cref{table:numbers} shows that \EHP is able to compete
within \numb{2.44x} of Intel MKL on both sparse and dense LA queries. 
On dense data, \EHP uses the attribute elimination optimization from 
\Cref{sec:query_optimization} to store dense annotations in single buffers
that are BLAS compatible and code generates to Intel MKL. Still, MKL  
produces only the output annotation, not the key values, so
\EHP incurs a minor performance penalty ($<$2\%) for producing the key values.
On sparse data, \EHP is able to compete with MKL when executing these LA queries
as pure aggregate-join queries. To do this, the attribute order 
and \texttt{GROUP BY} optimizations from \Cref{sec:query_optimization,sec:query_execution}
were essential. 
Although we tested more sophisticated optimizations, like cache 
blocking (by adding additional levels to the trie), we found that the performance
benefit of these optimizations did not consistently outweigh their added complexity.
In contrast, other relational designs fall flat on these LA queries. 
Namely, HyPer usually runs out of memory on the matrix multiplication query and, 
on the queries which it does complete, is often an order of magnitude slower 
than Intel MKL. Similarly, LogicBlox and MonetDB are at least an 
order of magnitude slower than Intel MKL on these queries.

\begin{table}
  \scriptsize
  \begin{center}
    \setlength{\tabcolsep}{3pt}
    \begin{tabular}{@{}ll|crrr@{}}
    \toprule
    Query &LH   &-Attr. Elim.   &-Sel.   &-Attr. Ord.  & -Group By \\
    \midrule
    Q1  &145ms        &2.54x  &-  &-        &875x \\
    Q3  &111ms    &2.46x &1.55x   &1.17x    &1.37x \\
    Q5  &128ms    &1.46x &1.80x   &72.68x   &1.12x \\
    Q6  &51ms        &4.82x &-     &-       &- \\
    Q8  &78ms     &-  &2.67x    &8815x      &13.49x \\
    Q9  &115ms      &- &2.54x   &18.96x     &1.13x \\
    Q10 &246ms      &1.70x &0.88x  &8.32x   &2.44x \\
    \bottomrule
    \end{tabular}
    \caption{Runtime for \EHP (LH) and relative performance without 
    optimizations on TPC-H queries. Run at scale factor
    10. `-' indicates there is no effect on the query.}
    \label{table:tpch_micros}
    \vspace{-5mm}
  \end{center}
\end{table}

\begin{table}
  \scriptsize
  \begin{center}
    \setlength{\tabcolsep}{3pt}
    \begin{tabular}{@{}lll|rrr@{}}
    \toprule
    Dataset &Query &LH &Attr. Elim. &-Attr. Order &-Group By \\
    \midrule
    hv15r   &SMV   &165ms   &-     &- &- \\
    hv15r   &SMM   &25.44s  &-     &oom &3.02x \\ 
    nlp240  &SMV   &171ms   &-     &- &- \\
    nlp240  &SMM   &4.69s   &-     &oom &185x \\ 
    16384   &DMV    &13.50ms &1.96x &- &- \\ 
    16384   &DMM    &4.43s   &500x  &- &- \\ 
    \bottomrule
    \end{tabular}
    \caption{Runtime for \EHP (LH) and relative performance without optimizations on linear algebra queries. `-' indicates that the optimization had no effect on the query.}  
    \label{table:la_micros}
    \vspace{-5mm}
    \end{center}
\end{table}

\subsection{Micro-Benchmarking Results}
\label{sec:micros}
We break down the performance impact of each optimization presented in
\Cref{sec:query_optimization,sec:cost_based,sec:query_execution}.

\paragraph*{Attribute Elimination} \Cref{table:tpch_micros,table:la_micros} show
that attribute elimination can enable up to a \numb{4.82x} performance advantage
on the TPC-H queries and up to a \numb{500x} performance advantage on 
dense LA queries. Attribute elimination is crucial 
on most on TPC-H queries, as these queries typically touch a small number of 
attributes from schemas with many attributes. Unsurprisingly, 
\Cref{table:tpch_micros} shows that attribute elimination provides the largest 
benefit on the scan TPC-H queries (1 and 6) because it allows \EHP to scan less 
data. On dense LA queries, \EHP calls Intel MKL with little overhead because
attribute elimination enables us to store each 
dense annotation in a BLAS acceptable buffer. As \Cref{table:numbers} shows, 
this yields up to a \numb{500x} speedup over processing these queries purely
in \EHP. 

\paragraph*{Selections} As shown in \Cref{table:tpch_micros}, pushing down 
selections can provide up to \numb{2.67x} performance advantage 
on the TPC-H queries we benchmark.
Although pushing down selections is generally useful, on
TPC-H Q10 it actually hurts execution time. 
This is because this optimization adds additional GHD nodes (intermediate 
results) to our query plans (something pipelining would fix). Still, 
the performance impact is small on this query ($<$12\%) and on average 
this optimization provided a 90\% performance gain across TPC-H queries.

\paragraph*{Attribute Order}
 As shown in \Cref{table:tpch_micros,table:la_micros}, the cost-based attribute ordering optimizer 
 presented in \Cref{sec:cost_based} can enable up to a \numb{8815x} performance 
 advantage on TPC-H queries and enables \EHP to run sparse matrix multiplication
 as a join query without running out of memory.
\Cref{table:tpch_micros,table:la_micros} show the difference between the 
best-cost and the worst-cost attribute orders. 
The most interesting queries here are TPC-H query 5 and TPC-H query 8. On TPC-H
query 5, it is essential that
the high cardinality \texttt{orderkey} attribute appears first. On TPC-H query
8, it is essential that the \texttt{partkey} attribute, which
was connected to an equality selection, appears first. The process of assigning
weights to the intersection costs in \Cref{sec:weights} ensures that orders 
satisfying these constraints are chosen. Finally, the cost-based 
attribute ordering optimizer is also crucial on sparse matrix multiplication. 
Here it is essential that the lower cost attribute order, with a projected 
away attribute before one that is materialized, is selected. This
order not only prevents a high cost intersection, but eliminates the 
computation and materialization of annotation values do not participate 
in the output (due to sparsity).

\paragraph*{GROUP BY} Finally, we show in \Cref{table:tpch_micros,table:la_micros} 
that \EHP's \texttt{GROUP BY} optimizers provide up to a
\numb{875x} performance advantage on TPC-H queries and up to a \numb{185x} 
performance advantage on LA queries. The \texttt{GROUP BY} annotation optimizer 
provides a \numb{875x} speedup on TPC-H query 1 because \texttt{GROUP BY} is the 
bottleneck operation in this query and our optimizer properly selects to use a 
per-thread hash map instead of libcuckoo here. The \texttt{GROUP BY} key optimizer 
provides a \numb{185x} speedup on SMM with the nlp240 dataset because
it correctly predicts that the many of the output key attribute sets are
sparse. As such \EHP's optimizer chooses to use a standard hash map to produce 
these sparse sets because using a bitset is highly inefficient (due to the 
amount of memory it wastes). This optimizer making the opposite choice provides 
a \numb{3x} performance advantage on SMM with the HV15R dataset. 


\begin{figure}
 \centering
 \includegraphics[width=0.9\linewidth]{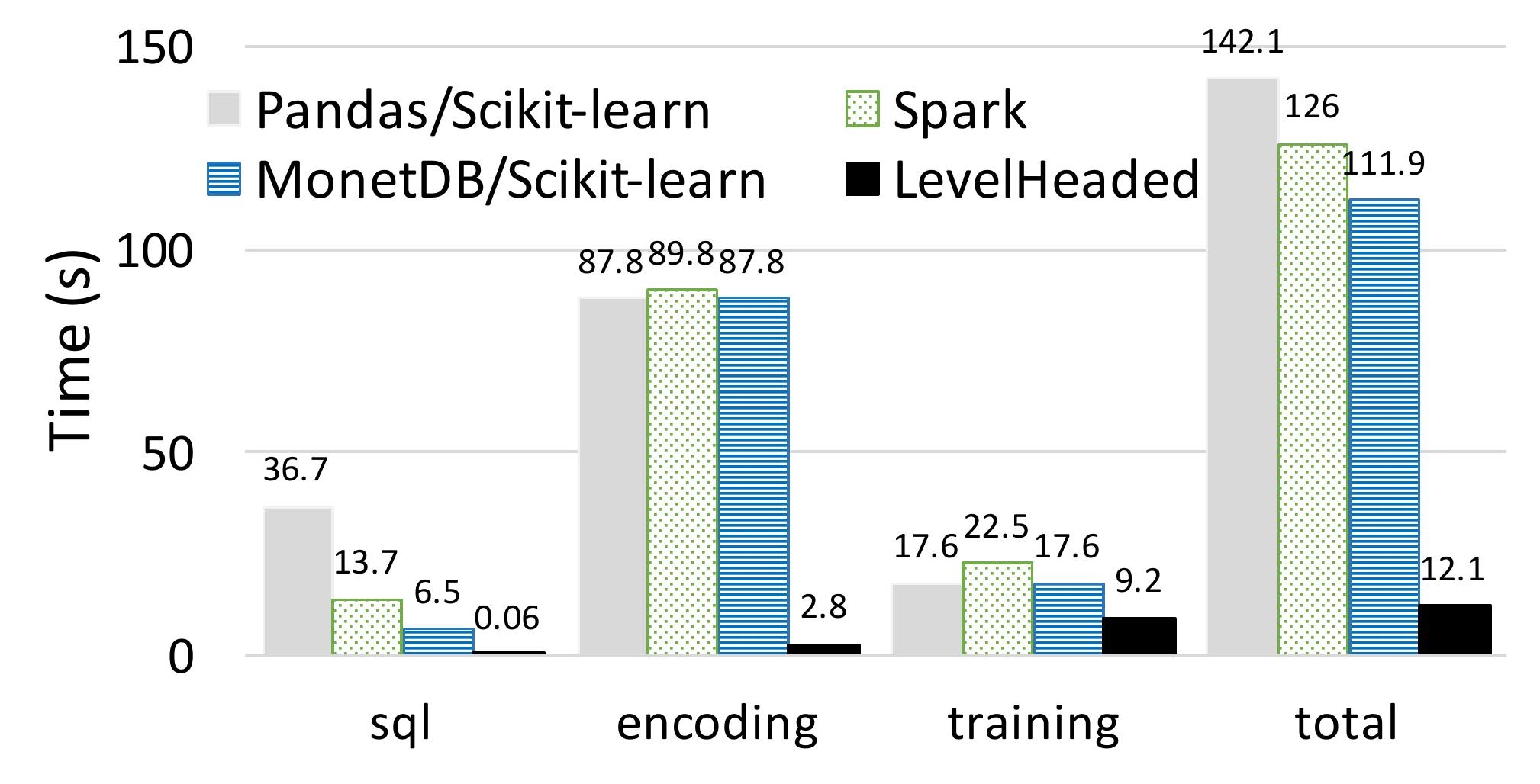}
\caption{Performance of engines on the voter classification application which
combines a SQL query, feature encoding, and the training of a machine learning 
model.}
\label{fig:mixed_chart}
\vspace{-2.5mm}
\end{figure}

\begin{table}
  \scriptsize
   \begin{center}
    \setlength{\tabcolsep}{10pt}
    \begin{tabular}{@{}lrrr@{}}
    \toprule
    Dataset & Conversion & SMV & Ratio \\
    \midrule
    Harbor &0.039s &0.0026s &15.00 \\
    HV15R &5.76s &0.17s  &33.88 \\
    nlp240 &7.11s &0.17s &41.82 \\
    \bottomrule
    \end{tabular}
    \caption{Runtime for dataset conversion, SMV query time in \EHP, and corresponding ratio (conversion/query). The conversion time measures Intel MKL's \texttt{mkl\_scsrcoo} library call 
    which is the (optimistic) time it takes to convert a column store to a acceptable sparse BLAS 
    format. The ratio is the number of times \EHP could run the query while a column store is converting the data. \label{fig:conv_tradeoff}}
    \label{fig:transfer_cost}
    \vspace{-5.5mm}
  \end{center}
\end{table}

\vspace{-2mm}
\section{Extension} \label{sec:extensions} 

We extend \EHP to show that such a unified query processing architecture
could enable faster end-to-end applications. 
To do this, we add
the ability for \EHP to process workloads that combine SQL queries and 
full machine learning algorithms (as described in \cite{abergerdemo}). With this, 
we show on a full-fledged application that \EHP can be an order of magnitude 
faster than the popular solutions 
of Spark v2.0.0, 
MonetDB Dec2016\footnote{\small Development build
with embedded Python \cite{monetdbnumpy}.}/Scikit-learn v0.17.1, and Pandas v0.18.1/Scikit-learn v0.17.1
(using the same experimental setting as \Cref{sec:experiments}).

\paragraph*{Application} We run a voter classification application 
\cite{monetdb} that joins and filters two tables to create a single feature
set which is then used to train a logistic regression model for 
five iterations. This application is a pipeline 
workload that consists of three pipeline phases: (1) a SQL-processing phase, (2) a feature
engineering phase where categorical variables are encoded, and (3) a 
machine learning phase. The dataset \cite{monetdb} consists of two tables: (1) one with
information such as gender and age for  
7,503,555 voters and
(2) one with information about the 2,751 precincts that the voters were registered in.

\paragraph*{Performance} \Cref{fig:mixed_chart} shows that \EHP
outperforms
Spark, MonetDB, and Pandas on the voter classification application by up to 
an order of magnitude. This is largely due to \EHP's optimized shared-memory
SQL processing and ability to minimize data transformations between 
the SQL and training phase. To expand on the cost of data transformations a bit further, 
in \Cref{fig:transfer_cost} we show the cost of 
converting from a column store to the compressed sparse row  
format used by most sparse library packages. This transformation is not 
necessary in \EHP as it always uses a single, trie-based data structure. 
As a result, \Cref{fig:transfer_cost} shows that up to 41 SMV queries can be
run in \EHP in the time that it takes for a 
column store to convert the data to a BLAS compatible format.  
Similarly, on the voter classification application LevelHeaded avoids expensive 
data transformations (in the encoding phase) by using its trie-based data 
structure for all phases.

\vspace{-2mm}
\section{Conclusions}

This paper introduced the \EHP engine and
demonstrated that a query architecture built around WCOJs can compete on 
standard BI and LA benchmarks. We showed that \EHP outperforms other relational 
engines by at least \numb{an order of magnitude} on LA queries, while remaining 
on average within \numb{31\%} of best-of-the-breed solutions on BI and LA 
benchmark queries. Our results are promising and suggest that 
such a query architecture could serve as the foundation for future 
unified query engines. 

\vspace{1.5mm}
{
\scriptsize
\textbf{Acknowledgments: } We thank LogicBlox for their helpful conversations 
and assistance in setting up our comparisons.
}
\vspace{0.2mm}

{
\small
\bibliographystyle{plain}
\bibliography{references}

\begin{thebibliography}{10}

\bibitem{intelmkl}
{Intel Math Kernel Library}.
\newblock \url{https://software.intel.com/en-us/mkl}.

\bibitem{utnla}
Oracle corporation.
\newblock \url{https://docs.oracle.com/cd/B1930- 6_01/index.html}.

\bibitem{abergerdemo}
Christopher Aberger, Andrew Lamb, Kunle Olukotun, and Christopher R{\'e}.
\newblock Mind the gap: Bridging multi-domain workloads with emptyheaded.
\newblock {\em Proceedings of the VLDB Endowment}, 10(12), 2017.

\bibitem{aberger2015emptyheaded}
Christopher~R Aberger, Susan Tu, Kunle Olukotun, and Christopher R{\'e}.
\newblock Emptyheaded: A relational engine for graph processing.
\newblock In {\em Proceedings of the 2016 International Conference on
  Management of Data}, pages 431--446. ACM, 2016.

\bibitem{aberger2016old}
Christopher~R Aberger, Susan Tu, Kunle Olukotun, and Christopher R{\'e}.
\newblock Old techniques for new join algorithms: A case study in rdf
  processing.
\newblock In {\em Data Engineering Workshops (ICDEW), 2016 IEEE 32nd
  International Conference on}, pages 97--102. IEEE, 2016.

\bibitem{khamis2015sf}
Mahmoud Abo~Khamis, Hung~Q Ngo, and Atri Rudra.
\newblock Faq: questions asked frequently.
\newblock In {\em Proceedings of the 35th ACM SIGMOD-SIGACT-SIGAI Symposium on
  Principles of Database Systems}, pages 13--28. ACM, 2016.

\bibitem{anderson1999lapack}
Edward Anderson, Zhaojun Bai, Christian Bischof, L~Susan Blackford, James
  Demmel, Jack Dongarra, Jeremy Du~Croz, Anne Greenbaum, Sven Hammarling, Alan
  McKenney, et~al.
\newblock {\em LAPACK Users' guide}.
\newblock SIAM, 1999.

\bibitem{aref2015design}
Molham Aref, Balder ten Cate, Todd~J Green, Benny Kimelfeld, Dan Olteanu, Emir
  Pasalic, Todd~L Veldhuizen, and Geoffrey Washburn.
\newblock Design and implementation of the logicblox system.
\newblock In {\em Proceedings of the 2015 ACM SIGMOD International Conference
  on Management of Data}, pages 1371--1382. ACM, 2015.

\bibitem{armbrust2015spark}
Michael Armbrust, Reynold~S Xin, Cheng Lian, Yin Huai, Davies Liu, Joseph~K
  Bradley, Xiangrui Meng, Tomer Kaftan, Michael~J Franklin, Ali Ghodsi, et~al.
\newblock Spark sql: Relational data processing in spark.
\newblock In {\em Proceedings of the 2015 ACM SIGMOD International Conference
  on Management of Data}, pages 1383--1394. ACM, 2015.

\bibitem{astrahan1976system}
Morton~M. Astrahan, Mike~W. Blasgen, Donald~D. Chamberlin, Kapali~P. Eswaran,
  Jim~N Gray, Patricia~P. Griffiths, W~Frank King, Raymond~A. Lorie, Paul~R.
  McJones, James~W. Mehl, et~al.
\newblock System r: relational approach to database management.
\newblock {\em ACM Transactions on Database Systems (TODS)}, 1(2):97--137,
  1976.

\bibitem{blackford2002updated}
L~Susan Blackford, Antoine Petitet, Roldan Pozo, Karin Remington, R~Clint
  Whaley, James Demmel, Jack Dongarra, Iain Duff, Sven Hammarling, Greg Henry,
  et~al.
\newblock An updated set of basic linear algebra subprograms (blas).
\newblock {\em ACM Transactions on Mathematical Software}, 28(2):135--151,
  2002.

\bibitem{brown2010overview}
Paul~G Brown.
\newblock Overview of scidb: large scale array storage, processing and
  analysis.
\newblock In {\em Proceedings of the 2010 ACM SIGMOD International Conference
  on Management of data}, pages 963--968. ACM, 2010.

\bibitem{intelbuzz}
Diane Bryant.
\newblock Live from intel ai day 2016.

\bibitem{corbett2013spanner}
James~C Corbett, Jeffrey Dean, Michael Epstein, Andrew Fikes, Christopher
  Frost, Jeffrey~John Furman, Sanjay Ghemawat, Andrey Gubarev, Christopher
  Heiser, Peter Hochschild, et~al.
\newblock Spanner: Google’s globally distributed database.
\newblock {\em ACM Transactions on Computer Systems (TOCS)}, 31(3):8, 2013.

\bibitem{davis2011university}
Timothy~A Davis and Yifan Hu.
\newblock The university of florida sparse matrix collection.
\newblock {\em ACM Transactions on Mathematical Software (TOMS)}, 38(1):1,
  2011.

\bibitem{delaney2000inside}
Kalen Delaney.
\newblock {\em Inside Microsoft SQL Server 2000}.
\newblock Microsoft Press, 2000.

\bibitem{duggan2015bigdawg}
Jennie Duggan, Aaron~J Elmore, Michael Stonebraker, Magda Balazinska, Bill
  Howe, Jeremy Kepner, Sam Madden, David Maier, Tim Mattson, and Stan Zdonik.
\newblock The bigdawg polystore system.
\newblock {\em ACM Sigmod Record}, 44(2):11--16, 2015.

\bibitem{garcia2008database}
Hector Garcia-Molina.
\newblock {\em Database systems: the complete book}.
\newblock Pearson Education India, 2008.

\bibitem{gottlob2005hypertree}
Georg Gottlob, Martin Grohe, Nysret Musliu, Marko Samer, and Francesco
  Scarcello.
\newblock Hypertree decompositions: Structure, algorithms, and applications.

\bibitem{gupta2014mesa}
Ashish Gupta, Fan Yang, Jason Govig, Adam Kirsch, Kelvin Chan, Kevin Lai, Shuo
  Wu, Sandeep~Govind Dhoot, Abhilash~Rajesh Kumar, Ankur Agiwal, et~al.
\newblock Mesa: Geo-replicated, near real-time, scalable data warehousing.
\newblock {\em Proceedings of the VLDB Endowment}, 7(12):1259--1270, 2014.

\bibitem{hellerstein2012madlib}
Joseph~M Hellerstein, Christoper R{\'e}, Florian Schoppmann, Daisy~Zhe Wang,
  Eugene Fratkin, Aleksander Gorajek, Kee~Siong Ng, Caleb Welton, Xixuan Feng,
  Kun Li, et~al.
\newblock The madlib analytics library: or mad skills, the sql.
\newblock {\em Proceedings of the VLDB Endowment}, 5(12):1700--1711, 2012.

\bibitem{idreos2012monetdb}
Stratos Idreos, Fabian Groffen, Niels Nes, Stefan Manegold, Sjoerd Mullender,
  Martin Kersten, et~al.
\newblock Monetdb: Two decades of research in column-oriented database
  architectures.
\newblock {\em A Quarterly Bulletin of the IEEE Computer Society Technical
  Committee on Database Engineering}, 35(1):40--45, 2012.

\bibitem{joglekar2016ajar}
Manas~R Joglekar, Rohan Puttagunta, and Christopher R{\'e}.
\newblock Ajar: Aggregations and joins over annotated relations.
\newblock In {\em Proceedings of the 35th ACM SIGMOD-SIGACT-SIGAI Symposium on
  Principles of Database Systems}, pages 91--106. ACM, 2016.

\bibitem{kemper2011hyper}
Alfons Kemper and Thomas Neumann.
\newblock Hyper: A hybrid oltp\&olap main memory database system based on
  virtual memory snapshots.
\newblock In {\em Data Engineering (ICDE), 2011 IEEE 27th International
  Conference on}, pages 195--206. IEEE, 2011.

\bibitem{kernert2015bringing}
David Kernert, Frank K{\"o}hler, and Wolfgang Lehner.
\newblock Bringing linear algebra objects to life in a column-oriented
  in-memory database.
\newblock In {\em In Memory Data Management and Analysis}, pages 44--55.
  Springer, 2015.

\bibitem{kumar2016model}
Arun Kumar, Robert McCann, Jeffrey Naughton, and Jignesh~M Patel.
\newblock Model selection management systems: The next frontier of advanced
  analytics.
\newblock {\em ACM SIGMOD Record}, 44(4):17--22, 2016.

\bibitem{kumar2016join}
Arun Kumar, Jeffrey Naughton, Jignesh~M Patel, and Xiaojin Zhu.
\newblock To join or not to join?: Thinking twice about joins before feature
  selection.
\newblock In {\em Proceedings of the 2016 International Conference on
  Management of Data}, pages 19--34. ACM, 2016.

\bibitem{li2014algorithmic}
Xiaozhou Li, David~G Andersen, Michael Kaminsky, and Michael~J Freedman.
\newblock Algorithmic improvements for fast concurrent cuckoo hashing.
\newblock In {\em Proceedings of the Ninth European Conference on Computer
  Systems}, page~27. ACM, 2014.

\bibitem{luo2017scalable}
Shangyu Luo, Zekai~J Gao, Michael Gubanov, Luis~L Perez, and Christopher
  Jermaine.
\newblock Scalable linear algebra on a relational database system.
\newblock In {\em Data Engineering (ICDE), 2017 IEEE 33rd International
  Conference on}, pages 523--534. IEEE, 2017.

\bibitem{mckinney2011pandas}
Wes McKinney.
\newblock pandas: a foundational python library for data analysis and
  statistics.
\newblock {\em Python for High Performance and Scientific Computing}, pages
  1--9, 2011.

\bibitem{meng2016mllib}
Xiangrui Meng, Joseph Bradley, Burak Yavuz, Evan Sparks, Shivaram Venkataraman,
  Davies Liu, Jeremy Freeman, DB~Tsai, Manish Amde, Sean Owen, et~al.
\newblock Mllib: Machine learning in apache spark.
\newblock {\em The Journal of Machine Learning Research}, 17(1):1235--1241,
  2016.

\bibitem{ngo2012worst}
Hung~Q Ngo, Ely Porat, Christopher R{\'e}, and Atri Rudra.
\newblock Worst-case optimal join algorithms.
\newblock In {\em Proceedings of the 31st ACM SIGMOD-SIGACT-SIGAI symposium on
  Principles of Database Systems}, pages 37--48. ACM, 2012.

\bibitem{ngo2014skew}
Hung~Q Ngo, Christopher R{\'e}, and Atri Rudra.
\newblock Skew strikes back: New developments in the theory of join algorithms.
\newblock {\em ACM SIGMOD Record}, 42(4):5--16, 2014.

\bibitem{patwary2015parallel}
Md~Mostofa~Ali Patwary, Nadathur~Rajagopalan Satish, Narayanan Sundaram,
  Jongsoo Park, Michael~J Anderson, Satya~Gautam Vadlamudi, Dipankar Das,
  Sergey~G Pudov, Vadim~O Pirogov, and Pradeep Dubey.
\newblock Parallel efficient sparse matrix-matrix multiplication on multicore
  platforms.
\newblock In {\em International Conference on High Performance Computing},
  pages 48--57. Springer, 2015.

\bibitem{pedregosa2011scikit}
Fabian Pedregosa, Ga{\"e}l Varoquaux, Alexandre Gramfort, Vincent Michel,
  Bertrand Thirion, Olivier Grisel, Mathieu Blondel, Peter Prettenhofer, Ron
  Weiss, Vincent Dubourg, et~al.
\newblock Scikit-learn: Machine learning in python.
\newblock {\em Journal of Machine Learning Research}, 12(Oct):2825--2830, 2011.

\bibitem{monetdbnumpy}
Mark Raasveldt.
\newblock Embedded python/numpy in monetdb.
\newblock \url{https://www.monetdb.org/blog/embedded-pythonnumpy-monetdb},
  2016.

\bibitem{monetdb}
Mark Raasveldt.
\newblock Voter classification using monetdb/python.
\newblock
  \url{https://www.monetdb.org/blog/voter-classification-using-monetdbpython},
  2016.

\bibitem{raman2013db2}
Vijayshankar Raman, Gopi Attaluri, Ronald Barber, Naresh Chainani, David
  Kalmuk, Vincent KulandaiSamy, Jens Leenstra, Sam Lightstone, Shaorong Liu,
  Guy~M Lohman, et~al.
\newblock Db2 with blu acceleration: So much more than just a column store.
\newblock {\em Proceedings of the VLDB Endowment}, 6(11):1080--1091, 2013.

\bibitem{schenk2008inertia}
Olaf Schenk, Andreas W{\"a}chter, and Martin Weiser.
\newblock Inertia-revealing preconditioning for large-scale nonconvex
  constrained optimization.
\newblock {\em SIAM Journal on Scientific Computing}, 31(2):939--960, 2008.

\bibitem{schiefer1998method}
B.~Schiefer.
\newblock Method for estimating cardinalities for query processing in a
  relational database management system, June~2 1998.
\newblock US Patent 5,761,653.

\bibitem{shute2013f1}
Jeff Shute, Radek Vingralek, Bart Samwel, Ben Handy, Chad Whipkey, Eric
  Rollins, Mircea Oancea, Kyle Littlefield, David Menestrina, Stephan Ellner,
  et~al.
\newblock F1: A distributed sql database that scales.
\newblock {\em Proceedings of the VLDB Endowment}, 6(11):1068--1079, 2013.

\bibitem{smith2014anatomy}
Tyler~M Smith, Robert Van De~Geijn, Mikhail Smelyanskiy, Jeff~R Hammond, and
  Field~G Van~Zee.
\newblock Anatomy of high-performance many-threaded matrix multiplication.
\newblock In {\em Parallel and Distributed Processing Symposium, 2014 IEEE 28th
  International}, pages 1049--1059. IEEE, 2014.

\bibitem{stonebraker2005c}
Mike Stonebraker, Daniel~J Abadi, Adam Batkin, Xuedong Chen, Mitch Cherniack,
  Miguel Ferreira, Edmond Lau, Amerson Lin, Sam Madden, Elizabeth O'Neil,
  et~al.
\newblock C-store: a column-oriented dbms.
\newblock In {\em Proceedings of the 31st international conference on Very
  large data bases}, pages 553--564. VLDB Endowment, 2005.

\bibitem{veldhuizen2012leapfrog}
Todd~L Veldhuizen.
\newblock Leapfrog triejoin: A simple, worst-case optimal join algorithm.
\newblock {\em arXiv preprint arXiv:1210.0481}, 2012.

\bibitem{whaley1998automatically}
R~Clint Whaley and Jack~J Dongarra.
\newblock Automatically tuned linear algebra software.
\newblock In {\em Proceedings of the 1998 ACM/IEEE conference on
  Supercomputing}, pages 1--27. IEEE Computer Society, 1998.

\bibitem{zaharia2010spark}
Matei Zaharia, Mosharaf Chowdhury, Michael~J Franklin, Scott Shenker, and Ion
  Stoica.
\newblock Spark: Cluster computing with working sets.
\newblock {\em HotCloud}, 10(10-10):95, 2010.

\end{thebibliography}
}

\end{document}